\documentclass[aps,prb,longbibliography,twocolumn,amsmath,showpacs,showkeys,superscriptaddress,floatfix]{revtex4-1}

\setcitestyle{numbers,square}

\usepackage{amsmath}
\usepackage{amssymb}
\usepackage{multirow,bigdelim}
\usepackage{array}
\usepackage{cancel}
\usepackage{epsfig}
\usepackage{graphicx}
\usepackage{graphics}
\usepackage{url}
\usepackage[english]{babel}
\usepackage{dcolumn}
\usepackage{color}
\usepackage[squaren]{SIunits}
\usepackage{bbold}
\usepackage[T1]{fontenc}
\usepackage{float}
\usepackage{color}
\usepackage{xcolor}
\usepackage{multirow}
\usepackage{bigdelim}
\usepackage{bm}

\renewcommand{\vec}{\boldsymbol}
\newcommand{\Onlinecite}{\cite}
\newcommand{\HHdown}{\raisebox{-1mm}{\rotatebox{90}{$\Lleftarrow$}}}
\newcommand{\HHup}{\raisebox{-1mm}{\rotatebox{90}{$\Rrightarrow$}}}
\newcommand{\LHdown}{\raisebox{-1mm}{\rotatebox{90}{$\Leftarrow$}}}
\newcommand{\LHup}{\raisebox{-1mm}{\rotatebox{90}{$\Rightarrow$}}}
\newcommand{\edown}{\raisebox{-1mm}{\rotatebox{90}{$\leftarrow$}}}
\newcommand{\eup}{\raisebox{-1mm}{\rotatebox{90}{$\rightarrow$}}}

\newcolumntype{C}[1]{>{\centering}p{#1}} 

\definecolor{exe}{RGB}{0,0,0}
\definecolor{exo}{RGB}{255,140,0}
\definecolor{eye}{RGB}{140,0,0}
\definecolor{eyo}{RGB}{135,135,135}
\definecolor{eze}{RGB}{0,0,140}
\definecolor{ezo}{RGB}{0,135,255}

\definecolor{oxe}{RGB}{0,0,0}
\definecolor{oxo}{RGB}{255,140,0}
\definecolor{oye}{RGB}{140,0,0}
\definecolor{oyo}{RGB}{135,135,135}
\definecolor{oze}{RGB}{0,0,140}
\definecolor{ozo}{RGB}{0,135,255}

\definecolor{nix}{RGB}{255,255,255}

\begin{document}

\title{Selection rules for the excitation of quantum dots by spatially structured light beams - Application to the reconstruction of higher excited exciton wave functions}

\author{M.~Holtkemper}
\affiliation{Institut f\"ur Festk\"orpertheorie, Universit\"at M\"unster,
Wilhelm-Klemm-Str.~10, 48149 M\"unster, Germany}

\author{G.~F.~Quinteiro}
\affiliation{IMIT and Departamento de F\'isica, Universidad Nacional del Nordeste,
Campus Deodoro Roca, Corrientes, Argentina}

\author{D.~E.~Reiter}
\affiliation{Institut f\"ur Festk\"orpertheorie, Universit\"at M\"unster,
Wilhelm-Klemm-Str.~10, 48149 M\"unster, Germany}

\author{T.~Kuhn}
\affiliation{Institut f\"ur Festk\"orpertheorie, Universit\"at M\"unster,
Wilhelm-Klemm-Str.~10, 48149 M\"unster, Germany}

\date{\today}

\begin{abstract}

Spatially structured light fields applied to semiconductor quantum dots yield fundamentally different absorption spectra than homogeneous beams. In this paper, we theoretically discuss the resulting spectra for different light beams using a cylindrical multipole expansion. For the description of the quantum dots we employ a model based on the effective mass approximation including Coulomb and valence band mixing. The combination of a single spatially structured light beam and state mixing allows all exciton states in the quantum dot to become optically addressable. Furthermore, we demonstrate that the beams can be tailored such that single states are selectively excited, without the need of spectral separation. Using this selectivity, we propose a method to measure the exciton wave function of the quantum dot eigenstate. The measurement goes beyond electron density measurements by revealing the spatial phase information of the exciton wave function. Thereby polarization sensitive measurements are generalized by including the infinitely large spatial degree of freedom.

\end{abstract}



\maketitle

\section{Introduction}\label{sec:introduction}
Spatially structured light (SSL) fields feature strong gradients in the light intensity, in contrast to standard Gaussian beams which are often approximated by plane waves. SSL appears in near field setups \cite{shibata2017excitation}, wave guides \cite{balistreri2000local}, photonic crystals \cite{abashin2006near} or other cavities \cite{cui2006hemispherical,cai2012integrated,curto2013multipolar} as well as in the far field of freely propagating beams \cite{schmiegelow2016transfer,kerber2017reading,quinteiro2015formulation,quinteiro2019reexamination}, denoted, e.g., as Hermite-Gaussian, Laguerre-Gaussian or Bessel beam. The spatial structure of such fields provides some interesting features: It is possible to get around the diffraction limit \cite{singh2017particle,huang2009super}. The infinite degree of freedom defined for instance by the orbital angular momentum of the beam provides a much more powerful approach to carry (quantum-)information than the two-dimensional polarization \cite{kerber2017reading,mair2001entanglement,molina2007twisted}, thereby enableing hyperentanglement \cite{mair2001entanglement, vaziri2002experimental, barreiro2008beating}, state cloning \cite{nagali2009optimal} and highly increased data transfer rates in optical communication \cite{wang2012terabit, bozinovic2013terabit, richardson2013space}.

We will study theoretically the interaction of SSL with self-assembled semiconductor quantum dots (QDs). QDs provide a quantized and widely adjustable electronic level structure and are established components in many modern applications \cite{portalupi2019inas,dusanowski2019near,reiter2019distinctive,trivedi2020generation,kneissel2020semiconductor,kupko2020tools}, especially in the context of quantum information technology \cite{michler2017quantum,huber2018semiconductor,ramsay2010review,warburton2013single}. In addition to the energetically lowest four levels, i.e. the ground state bright and dark excitons, which play a central role for many applications, also higher excited states become interesting. The latter can be used, e.g., to describe metastable states in charged QDs \cite{hinz2018charge}, for multiexciton states \cite{hawrylak1999excitonic,akimov2002fine,chithrani2005electronic,babinski2006emission,arashida2011four,kuklinski2011tuning,molas2012fine,piketka2013photon,molas2016quadexciton}, for state preparation schemes \cite{schwartz2015deterministic,traum2019ultrafast}, to study dephasing and relaxation processes \cite{htoon2001carrier,huneke2011role,hinz2018charge} or for resonant absorption within a QD \cite{gammon1996fine,htoon2001carrier,benny2012excitation,molas2016energy,holtkemper2018influence}. To utilize higher excited electronic states, they need to be addressable, selectively excitable and identifiable. In this paper, we will show that these three prerequisites are highly improved when using SSL.\\
\begin{enumerate}
\item{Addressability: Plane wave selection rules allows only specific electronic transitions. The number of addressable states is highly increased when using SSL with their corresponding multipole transitions \cite{zurita2002multipolare,zurita2002multipolarm,quinteiro2009electronic}. However, several electronic states stay unaddressable within simplified QD models. We show within a realistic QD model including Coulomb interactions and valence band mixing, that each eigenstate becomes accessible by an appropriate SSL field. The oscillator strength of these previously dark states varies from negligible to strong, depending on the individual state mixtures. We discuss relevant coupling mechanisms and the influence of symmetry breaking. The oscillator strengths are visualized within calculated absorption spectra.}
\item{Selectivity: The QD's eigenstates are often energetically close and not individually addressable by short and thus broadband laser pulses. This limits for instance the temporal resolution in pump probe experiments utilizing higher excited states \cite{hinz2018charge}. One known way around this limit is a polarization sensitive excitation, where even energetically arbitrarily close states (like the horizontally and vertically polarized exciton ground states) can be addressed selectively. However, the polarization sensitive excitation is limited to the two-dimensional spin degree of freedom. With our scheme we show that using the infinite spatial degree of freedom of SSL in additon to the polaritation, one can highly increase the possiblity of selective excitation.}
\item{Identification: To identify an electronic eigenstate within a QD, one has to measure its wave function. The measurement of electron densities in QDs is possible, e.g., by scanning tunneling \cite{maltezopoulos2003wave} or magnetotunneling spectroscopy \cite{vdovin2000imaging}. We propose a method to reconstruct the wave function $\Psi^{\text{exciton}}(\vec{r}_{\text{hole}},\vec{r}_{\text{elec.}})$ of excitons for $\vec{r}_{\text{hole}}=\vec{r}_{\text{elec.}}$ from pure optical experiments. Our method goes beyond today's measurements, since not just electronic densities $|\Psi^{\text{exciton}}|^2$ are measurable, but the wave function $\Psi^{\text{exciton}}$ itself (of course except for a global phase). Our proposed method is not restricted by the diffraction limit.}
\end{enumerate}

The paper is structured as follows: The model for the QD and the light-matter-interaction is given in Sec.~\ref{sec:model}. Analytical selection rules and absorption spectra are presented for a simplified QD model in Sec.~\ref{sec:reducedmodelab} and are then generalized to our full model in Sec.~\ref{sec:coulombinteractionvbm}. A proposal to measure the wave functions of the QD's eigenstates is given in Sec.~\ref{sec:dipolepolar}. Section~\ref{sec:realis} discusses the experimental viability of the considered light modes. Concluding statements are given in Sec.~\ref{sec:conclusion}. Finally, the appendices \ref{sec:appeqalpr}-\ref{sec:appBesToNod} provide some background information and additional details.

\section{Model}\label{sec:model}

\subsection{QD Model}\label{sec:qdmodel}
We model the electronic level structure of a QD based on the envelope function approximation. In this approach, the single particle wave functions are separated into a Bloch and an envelope part.\\
The Bloch part is expanded within a basis containing the $\Gamma$-point states of the heavy hole (HH), light hole (LH) and lowest conduction (EL) bands, described by their (pseudo-)spins $\pm\frac{3}{2}$ (or $\,\HHup$ and $\HHdown$), $\pm\frac{1}{2}$ ($\Uparrow$ and $\Downarrow$) and $\pm\frac{1}{2}$ ($\uparrow$ and $\downarrow$), respectively. We use the $z$-axis (which is the growth direction of the QD) as the quantization axis.
For excitons, we get eight possible combinations of the spin states. Since we assume - as it is realistic - a broken cylindrical symmetry of the QD, the most convenient basis is defined by the linearly polarized HH exciton states $\epsilon_{x}$, $\epsilon_{y}$, $\epsilon_{{z}}$ and $\epsilon_{{{0}}}$ as well as the LH exciton states $\epsilon_{\tilde{x}}$, $\epsilon_{\tilde{y}}$, $\epsilon_{\tilde{z}}$ and $\epsilon_{\tilde{0}}$, which are listed in Tab.~\ref{polarselection}. We use a phase convention as in Ref.~\cite{holtkemper2018influence}.\\
The envelope functions are expanded in terms of Cartesian Hermite-Gaussian functions $\Phi_{\vec{a}}(\vec{r})=\tilde{\Phi}_{a^x}(x)\tilde{\Phi}_{a^y}(y)\tilde{\Phi}_{a^z}(z)$ with quantum numbers $\vec{a}=(a^x,a^y,a^z)$. States with $a^x+a^y+a^z=0,1,2...$ will be called $s$, $p$, $d$ ...-like states. LH states are labeled by capital letters $S$, $P$, $D$ ... . If necessary, indices provide a distinction between $a^x$, $a^y$ and $a^z$, e.g. $d_{xy}$ for $\vec{a}=(1,1,0)$. Since excitations in different in-plane direction are often similar, we use ``$\text{inpl.}$'' as a label for any in-plane direction, thus for example $d_{\text{inpl.}}$ is a shortcut for $d_{xx}$, $d_{xy}$ and $d_{yy}$ states. Transitions as well as the associated exciton states are labeled in the scheme hole~$\to$~electron.\\
A full configuration interaction (CI) approach is used to account for correlation effects. The CI-basis states are given by electron-hole product states.
Our Hamiltonian reads
$$\hat{H} = \hat{H}_\text{EMA} + \hat{H}_\text{DCI} + \hat{H}_\text{SRE} + \hat{H}_\text{VBM}$$
and includes the QD confinement within an effective mass approximation (EMA), the direct (DCI) and short range exchange (SRE) Coulomb interactions as well as valence band mixing via the offdiagonal elements of a four-band Luttinger model (VBM).
As an approximation to the QD confinement, we use an anisotropic harmonic potential treated in Cartesian coordinates. The frequencies of the potential $\omega_{b,\alpha}=\frac{4 \hbar}{m_{b,\alpha} \beta^2_{b} L^2_{\alpha}}$ ($b$ denoting the band index, $\alpha$ the direction) are chosen such that the probability density of the ground state is reduced to $\frac{1}{e}$ at the distance $\pm\frac{1}{2}L_\alpha$ from the QD center. The QD diameters $L_{\alpha}$ are fixed to ($5.8\times5.0\times2.0$) nm$^3$, representing the flat geometry with slightly broken cylindrical symmetry of a typical self assembled QD. The wave functions of the holes are assumed to be broader than those of the electrons by a factor $\beta_\text{HH/LH}=\beta=1.15$ (we set $\beta_\text{EL}=1.0$). We use the material parameters of CdSe \cite{ekimov1993absorption, laheld1997excitons}, with the band gap of 1840~meV and the effective masses $m_{b,\alpha}$ (in terms of the free-space electron mass $m_0$): $m_\text{EL}=0.13m_0$, $m_{\text{HH},x/y}\approx 0.38m_0$, $m_{\text{HH},z}=m_0$, $m_{\text{LH},x/y}\approx 0.65m_0$ and $m_{\text{LH},z}\approx 0.31m_0$, deduced from the Luttinger parameters $\gamma_1=2.1$, $\gamma_2=\gamma_3=0.55$. For the DCI, we use the static dielectric constant of bulk CdSe \cite{laheld1997excitons} of $\epsilon_r=9.2$. The parameter for the coupling strength of SRE is set to $M_\text{SRE}=1.47$~meV by fitting experimental data from Ref.~\cite{hinz2018charge}.
Details of the model are discussed in Ref.~\Onlinecite{holtkemper2018influence}.

\subsection{Light matter interaction}\label{sec:lmi}

The light-matter interaction can be described in different gauges. The so called twisted-light gauge \cite{quinteiro2015formulation} is highly adapted to describe the interaction of higher multipole modes with matter. However, the use of the twisted-light gauge in conjunction with the envelope function approximation is theoretically challenging and requires a re-examination of the typical approximations involved in deriving the envelope function approximation and its correspondence with the approximations made on the vector and scalar potential. Such an analysis is out of the scope of the present article, and thus we will here make use of the standard minimal coupling Hamiltonian in Coulomb gauge to calculate the transition matrix elements. Accordingly, neglecting quadratic terms in the fields, we consider
\begin{align}
\hat{H}_{\gamma}(t)=\frac{e}{m_0} \vec{A}(\vec{r},t)\cdot \hat{\vec{p}}
\end{align}
with the elementary charge $e$, the vector potential $\vec{A}$ and the canonical momentum operator $\hat{\vec{p}}$. Following the typical approach, we assume monochromatic waves, use the rotating wave approximation and Fermi's golden rule to get an absorption $\sim |\langle \Psi_{a_1,b_1}|\hat{\tilde{H}}_{\gamma}| \Psi_{a_2,b_2} \rangle|^2 \cdot \delta(E_{X}-\hbar \omega)$ with the energy of the exciton $E_{X}$, the frequency of the light field $\omega$ and $\hat{\tilde{H}}_{\gamma}=\frac{e}{m_0} \tilde{\vec{A}}(\vec{r})\cdot \hat{\tilde{\vec{p}}}^+ + h.c.$ with the complex vector potential $\tilde{\vec{A}}$ defined by $\vec{A}(\vec{r},t) = \tilde{\vec{A}}(\vec{r})e^{-i\omega t} + h.c.$, matrix elements of $\hat{\tilde{\vec{p}}}$ just causing transitions from higher to lower energetic states and $h.c.$ the hermitian conjugate term. Each absorption line is widened by a Lorentzian function with a full width at half maximum of 0.1 meV to improve visibility. The matrix elements can be approximated by using the typical steps of the envelope function approximation: First, one separates the wave function  $\Psi_{a,b}(\vec{r})=\sqrt{V_\text{uc}}\,\Phi_a(\vec{r})\,u_b(\vec{r})$ into an envelope $\Phi_a$ and a Bloch function $u_b$, with $V_\text{uc}$ being the volume of the unit cell. Then the transition $\vec{r}\to \vec{r}'+\vec{R}$ with the position of the actual unit cell $\vec{R}$ and the relative position within this unit cell $\vec{r}'$ is applied. This is accompanied by $V_\text{uc} \sum_{\vec{R}} \int d^3r \to \int d^3R \int\limits_{V_\text{uc}} d^3r'$ and $\Phi(\vec{r}) = \Phi(\vec{R}+\vec{r}') \approx \Phi(\vec{R})$. Furthermore we use the approximation, that the field varies slowly over a single unit cell, thus we neglect in
\begin{align}
\label{eq:SVEA}
&\tilde{\vec{A}}(\vec{R}+\vec{r}') \cdot \hat{\vec{p}} =\tilde{\vec{A}}(\vec{R}) \cdot \hat{\vec{p}} + \sum_{\alpha \in \{x,y,z\} } \partial_{R_{\alpha}} \tilde{\vec{A}}(\vec{R}) \cdot r_{\alpha}\hat{\vec{p}} + ...
\end{align}
all but the zeroth order term. Since $\hat{\vec{p}}$ results in dipole moments on the atomic length scale of the Bloch functions (see App.~\ref{sec:appeqalpr}), higher order terms like $r_{\alpha} \hat{\vec{p}}$ would result in higher multipole transitions on the atomic scale, which can be neglected safely. However, with the full dependency of $\tilde{\vec{A}}(\vec{R})$ on $\vec{R}$, all higher multipole transitions on the mesoscopic scale of the envelopes are still included. In total we get

{\footnotesize
\begin{align}
\label{eq:LMWW}
&\langle \Psi_{a_1,b_1}|\hat{\tilde{H}}_{\gamma}| \Psi_{a_2,b_2} \rangle \notag\\
=&V_\text{uc}\int d^3r \,\, \Phi_{a_1}^*(\vec{r})\,u_{b_1}^*(\vec{r})\, \left[\frac{e}{m_0} \tilde{\vec{A}}(\vec{r})\cdot \hat{\tilde{\vec{p}}}^+ \right]\, \Phi_{a_2}(\vec{r})\,u_{b_2}(\vec{r}) +h.c. \notag\\
\approx & \underbrace{\int d^3R\, \, \Phi_{a_1}^*(\vec{R})\,\tilde{\vec{A}}(\vec{R})\,\Phi_{a_2}(\vec{R})}_{\vec{M}_{a_1,a_2}(\tilde{\vec{A}})} \notag\\
        & \cdot \underbrace{\int\limits_{V_\text{uc}} d^3r' \,\, u_{b_1}^*(\vec{r}')\, \left[e \frac{\hat{\tilde{\vec{p}}}'^+}{m_0} \right]\, u_{b_2}(\vec{r}')}_{\vec{\mu}_{b_1,b_2}} + \, h.c. \, .
\end{align}
}

The microscopic dipole moments $\vec{\mu}_{b_1,b_2}$ for the eight basis states are known \cite{bastard1990wave} up to a constant factor and listed in Tab.~\ref{polarselection}. 

\begin{figure}[H]
\includegraphics[width=1.0\columnwidth]{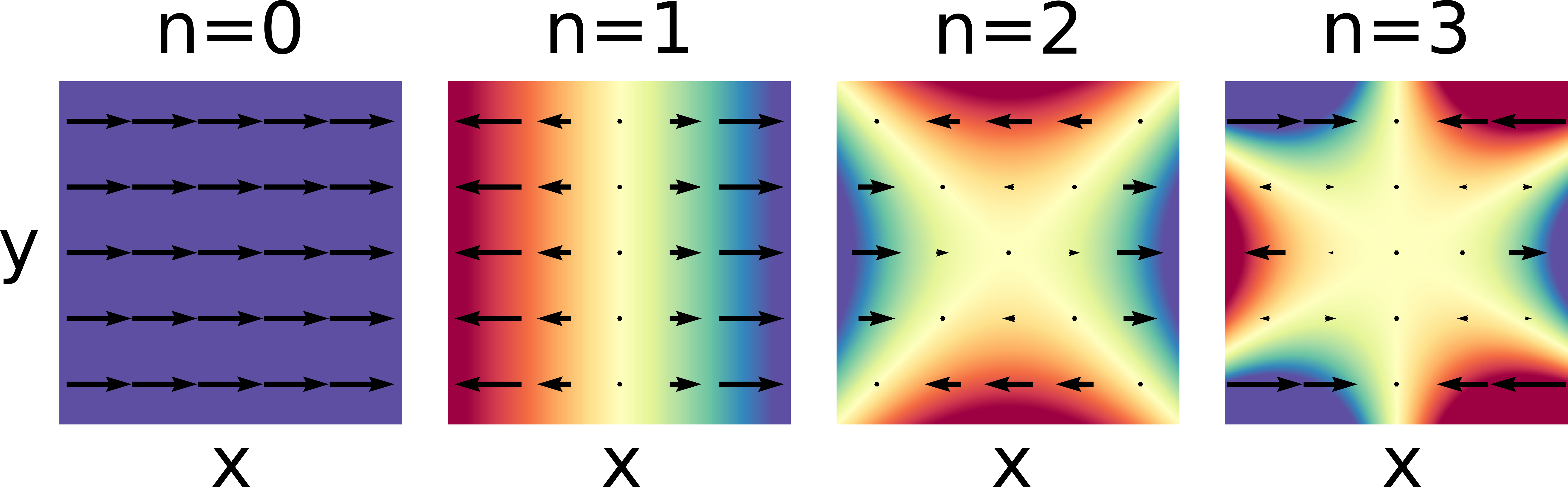}
\caption{Beam profiles as described in Eq.~\eqref{eq:ideallightE} for $\theta=0$, $\alpha=x$ and $n\in\{0,1,2,3\}$. Red/blue areas symbolize opposing orientations of the field.}\label{fig:idealLE}
\end{figure}

The focus of this work is the evaluation of the mesoscopic transition matrix elements $\vec{M}_{a_1,a_2}(\tilde{\vec{A}})$ for specific light fields. They depend on the spatial structure of the light field, which can be arbitrarily complicated. Typically, this problem is fixed by a spherical multipole expansion of the light fields, such that selection rules can be given in a compact form. However, we are interested in the interaction with spatially structured beams, whose cylindrical geometry can be used more directly within a cylindrical multipole expansion instead of the spherical one. In the cylindrical multipole expansion the radial dependency of an arbitrary beam profile is described by Bessel functions $J_n(q_r r)$ and the angle dependency either by the real functions $\cos(n\varphi)$ and $\sin(n\varphi)$ or by their complex counterparts $e^{\pm i n \varphi}$. We use the real representation, since in particular in the case of a broken cylindrical symmetry of the QD, it provides more specific selection rules. The index $n$ describes the order of the cylindrical multipole expansion, which is basically the number of nodal lines crossing the beam center (see Fig.~\ref{fig:idealLE}). If we assume the QD to be on the beam axis, we can use the approximation $J_n(q_r r) \sim r^n$ around the QD, since the beam profile is typically much larger than the QD i.e., $q_r^{-1}\gg L_{x/y}$. For linear polarizations along $\vec{e}_{\alpha}$ (with $\alpha\in \{x,y,z\}$), an arbitrary beam profile close to the QD can be described in the basis
\begin{align}
& \tilde{\vec{A}}_{n,\theta,\alpha}(\vec{r}) = A_{0} \left( \frac{r}{R_\text{QD}} \right)^n \cos(n \varphi-\theta) \, \vec{e}_{\alpha} \label{eq:ideallightE}
\end{align}
with $R_\text{QD}=\frac{1}{4}(L_x +L_y)$. $\theta$ is introduced to describe the cosine or sine function and can take the value $0$ or $\frac{\pi}{2}$. Typical geometries of those fields are plotted in Fig.~\ref{fig:idealLE} for $\vec{e}_x$ polarized light.

\begin{widetext}

\begin{table}[h]
$\phantom{aa}$\\[-3mm]
\centering
\begin{tabular}{c||c|c|c|c||c|c|c|c}
state & $\epsilon_{x}$& $\epsilon_{y}$&$\epsilon_{{z}}$& $\epsilon_{{{0}}}$& $\epsilon_{\tilde{x}}$& $\epsilon_{\tilde{y}}$& $\epsilon_{\tilde{z}}$ & $\epsilon_{\tilde{0}}$ \\
\hline
\hline
 & & & & & & & \\[-9pt]
spin configuration & $\frac{1}{\sqrt{2}} (\HHup \edown - \HHdown \eup)$& $\frac{-i}{\sqrt{2}} (\HHup \edown + \HHdown \eup)$& $\frac{1}{\sqrt{2}} (\HHup \eup + \HHdown \edown)$& $\frac{i}{\sqrt{2}} (\HHup \eup - \HHdown \edown)$& $\frac{1}{\sqrt{2}} (\LHup \eup - \LHdown \edown)$& $\frac{-i}{\sqrt{2}} (\LHup \eup + \LHdown \edown)$& $\frac{1}{\sqrt{2}} (\LHup \edown + \LHdown \eup)$ & $\frac{i}{\sqrt{2}} (\LHup \edown - \LHdown \eup)$ \\[2pt]
\cline{1-9}
 & & & & & & & \\[-9pt]
parity $[P^x,P^y,P^z]$ & $[o,e,e]$       & $[e,o,e]$       & $[e,e,o]$                 & $[o,o,o]$                         & $[o,e,e]$ & $[e,o,e]$ & $[e,e,o]$ & $[o,o,o]$ \\[0mm]
\cline{1-9}
 & & & & & & & \\[-9pt]
dipole moment $\frac{\vec{\mu}}{\mu_0}$ & $\vec{e}_{x}$       & $\vec{e}_{y}$       & $0$                 & $0$                         & $\sqrt{\frac{1}{3}}\vec{e}_{x}$ & $\sqrt{\frac{1}{3}}\vec{e}_{y}$ & $\sqrt{\frac{4}{3}}\vec{e}_{z}$ & $0$ \\
\end{tabular}
\caption{Definition of the basis functions used to describe the Bloch part of the wave function, and corresponding dipole moments. In the third row, $o$ and $e$ refer to even and odd parity, respectively.}
\label{polarselection}
\end{table}

\end{widetext}

\section{Reduced QD model}\label{sec:reducedmodelab}
\begin{widetext}

\begin{figure}[H]
\centering
\includegraphics[width=0.86\textwidth]{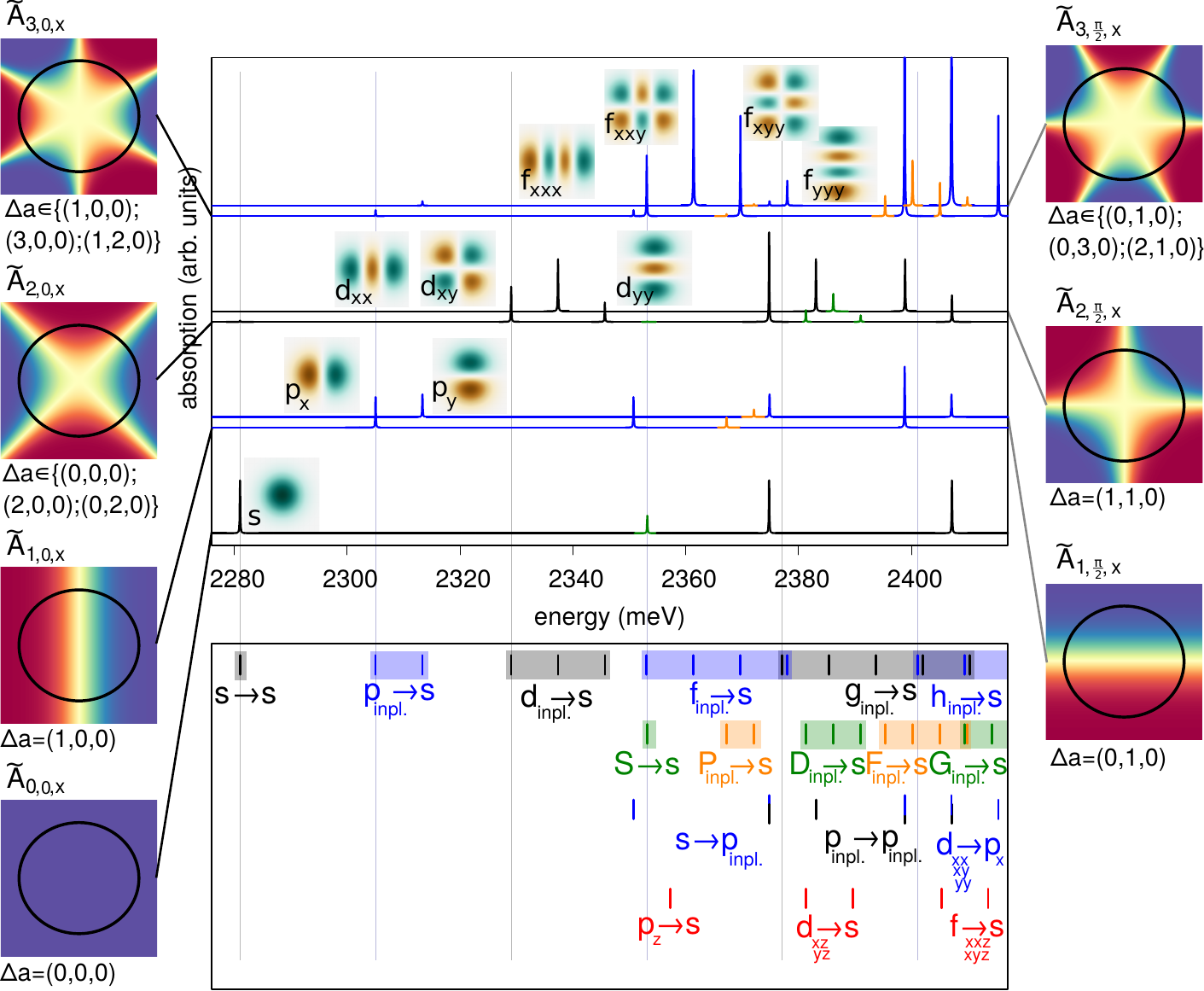}
\protect\caption{Absorption spectra for a QD in the simplified model ($\beta=1$ and without Coulomb interaction and VBM). Below, all existing exciton states are displayed at the corresponding transition energy. Absorption lines are marked in the corresponding colors as a guide for the eye. At the sides we sketch the geometry of the light field amplitudes around the QD in the $z=0$ plane as well as the selection rules in the form $\Delta a = (\Delta a^x, \Delta a^y, \Delta a^z)$. Blue/red mark oppositional phases of the field. A linear polarization in $x$-direction is assumed ($y$-polarization equivalent). Between the spectra, we sketch the envelopes of the lowest few single particle states in the $z=0$ plane.}\label{fig:RedMod}
\end{figure}

\end{widetext}

Before studying our full model, it is instructive to consider a simple envelope function approach with equal confinement lengths for electron and hole ($\beta=1$) and with uncoupled electron-hole pairs (without Coulomb interactions and VBM). In such a model, products of our single particle basis states $\Phi_{a} u_b$ already represent eigenstates of the system. The energetic structure of these states is given by labeled dashes in the lower panel of Fig.~\ref{fig:RedMod}. At low energies, we find mainly those excitons composed of an in-plane excited HH and an electron in the $s$-like conduction band state, e.g. $s\to s$, $p_{\text{inpl.}}\to s$, $d_{\text{inpl.}}\to s$, ... (see first row of dashes). Thereby states excited in different in-plane directions, like $p_x\to s$ and $p_y\to s$, form clusters in energy, since the diameter of the QD in $x$- and $y$-direction is similar. Because in these excitons, the electron is always in the isotropic $s$-like state, the geometry of the exciton's envelope is defined by the geometry of the hole state (which is sketched for the lowest few excitons between the spectra in Fig.~\ref{fig:RedMod}). Besides these HH excitons, at higher energies a similar level structure occurs for the LH excitons (second row of dashes). In addition, the electron can also be excited, resulting for example in $s\to p_{\text{inpl.}}$ or $p_{\text{inpl.}}\to p_{\text{inpl.}}$ excitons (third row of dashes). Because the QD is not a purely two-dimensional structure but has a finite height, we also get excitons excited in $z$-direction (fourth row of dashes). One should note, that each level (dash) has a four-fold spin degeneracy.

In Fig.~\ref{fig:RedMod}, we show absorption spectra for different cylindrical multipole modes (see Eq.~\eqref{eq:ideallightE}) for n=0,1,2,3. The spectra are displayed for $x$-polarization and are identical for $y$-polarization. Spectra for $z$-polarization can be deduced by setting the HH transitions to zero and upscaling the LH transitions by a factor of four (see Tab.~\ref{polarselection}). The geometries of the light field amplitudes in the $z=0$ plane are sketched on the left (right) hand side of the spectra for $\theta=0$ ($\theta=\frac{\pi}{2}$).

To understand the absorption patterns, we need to consider the transition matrix elements $\vec{M}_{a_1,a_2}(\tilde{\vec{A}})$ (see Eq.~\eqref{eq:LMWW}). These are basically given by the overlap between the amplitude distribution of the light field (given on the side of the spectra) and the envelope of the exciton (given for several lower excitons between the spectra). These overlaps and thereby the selection rules are easily estimated by a visual comparison. In particular the parity of light field amplitude and envelope need to be equal in each direction to allow for an absorption. As an example, the light field $\tilde{\vec{A}}_{1,0,x}$ is odd in $x$- and even in $y$-direction. This light field will just interact with excitons with the same parity, like the $p_x\to s$ exciton. The exciton $p_y\to s$, which is even in $x$- and odd in $y$-direction, has no overlap with $\tilde{\vec{A}}_{1,0,x}$ and accordingly no absorption line is visible. It turns out, that the difference between the envelope quantum number of the exciton's hole and electron, namely $\Delta a^{\alpha} = |a^{\alpha}_{\text{hole}}-a^{\alpha}_{\text{elec.}}|$, is a well suited quantity to describe selection rules. With our graphical approach, we can easily deduce the selection rule, that for an interaction with the field $\tilde{\vec{A}}_{1,0,x}$, the exciton has to fulfill $\Delta a^x=$ odd and $\Delta a^y=$ even. In fact, the selection rule can be further restricted to the explicit values $\Delta a=(\Delta a^x,\Delta a^y,\Delta a^z)=(1,0,0)$. Accordingly just absorption lines for $p_x\to s$, $s\to p_x$, $P_x\to s$ and $d_{xx}\to p_x$ are observed in the given energetic range. The corresponding envelope selection rules are given below the field profiles.
An analytical derivation of the envelope selection rules for arbitrary cylindrical multipole modes is given in App.~\ref{sec:appderanaselect}. Summarized, we get a light-matter interaction if:
\begin{enumerate}
\item{For $\theta=0$: The parity of $\Delta a^x$ has to be the same as the parity of $n$, while $\Delta a^y$ has to be even;\\
For $\theta=\frac{\pi}{2}$: The parity of $\Delta a^x$ has to be different from the parity of $n$, while $\Delta a^y$ has to be odd}
\item{$\sum_{\nu \in \{x,y\}} \Delta a^{\nu} \leq n$; $\Delta a^z = 0$}
\item{$\sum_{\nu \in \{x,y\}} \left(a_{\text{elec.}}^{\nu}+a_{\text{hole}}^{\nu}\right) \geq n$. This third rule just holds exactly for $L_x=L_y$. However, one could stretch the light field to match the same oval form as the QD to restore this selection rule.}
\end{enumerate}
The first two ``strong'' envelope selection rules are considered for the rules given in Fig.~\ref{fig:RedMod}.

After this general introduction to the selection rules, we highlight some details of the absorption spectra:

While $p_x\to s$ / $p_y\to s$ can be accessed separately by different light modes, for higher $\text{HH}_{\text{inpl.}}\to s$ envelope clusters no full selectivity between the different states is achieved. Here the parity determines which of the states are addressed by the same light field: For even $n$, all envelope states following the symmetry $\Delta a=$(odd,odd,even)=(o,o,e) or (e,e,e) are addressed together, and for odd $n$ the states of the form (o,e,e) or (e,o,e). The parity of some selected states is given in Tab.~\ref{envselection}. This ``reduced'' selectivity is a side effect of the fact, that the radial variation of the light field is restricted (see App.~\ref{sec:appderanaselect}, subsection on Hermite-Gaussian beam profiles).

In a cylindrically symmetric QD, the $\text{HH}_{\text{inpl.}}\to s$ transitions would give the energetically lowest possible transitions, caused by the third selection rule. This rule is slightly broken in the case of a broken cylindrical symmetry, causing for $n=2$ the small (in Fig.~\ref{fig:RedMod} hardly visible without zooming into the figure) $s\to s$ and $S\to s$ peaks and for $n=3$ the small $p_{\text{inpl.}}\to s$, $s\to p_{\text{inpl.}}$ and $P_{\text{inpl.}}\to s$ peaks.\\
In addition to the $\text{HH}_{\text{inpl.}}\to s$ transitions, there are the corresponding $\text{LH}_{\text{inpl.}}\to s$ transitions visible at higher energies. Furthermore, we find in the spectra:
\begin{itemize}
\item{For $n=0$ the $p_x\to p_x$ and $p_y\to p_y$ transitions.}
\item{For $n=1$ the $s\to p_x$/$s\to p_y$ and $d_{xx}\to p_x$/$d_{xy}\to p_x$ transitions.}
\item{For $n=2$ in addition to the already with $n=0$ accessible $p_x\to p_x$/$p_y\to p_y$ also the $p_y\to p_x$/$p_x\to p_y$ transitions.}
\item{For $n=3$ in addition to the already with $n=1$ accessible $d_{xx}\to p_x$/$d_{xy}\to p_x$ also the $d_{yy}\to p_x$ transition.}
\end{itemize}
Not reachable at this level of approximation are excitons with hole and electron in different excitation levels in $z$-direction ($\Delta a^z \neq 0$, marked in red). To enable such transitions, the light field needs to have a nodal line in $z$-direction which can be created, e.g., by a standing wave or an incidence of the beam from the side.
Furthermore, each exciton level still has a fourfold spin degeneracy, at this level of approximation. Therefore, spin selection rules, even if they exist, do not show up in the spectra. Spin selection rules will become apparent in the next section.

The intensities of transitions excited in different directions (e.g. $p_x\to s$ and $p_y\to s$) are not equal. Because the interaction of higher modes with the QD results from the finite light field in the outer regions of the QD, and the state excited in the direction of the wider QD confinement (here $p_x\to s$) has larger contributions in these outer regions, the light matter coupling is stronger for these states.

When we rotate the orientation of the nodal planes for a fixed value of $n$, we get a continuous change between the two plotted spectra. This implies, that one can determine the orientation of the QD by an alignment of the light field to the case, where the spectra are most selective.

\section{Full QD model}\label{sec:coulombinteractionvbm}
\begin{widetext}

\begin{figure}[H]
\centering
\includegraphics[width=0.9\textwidth]{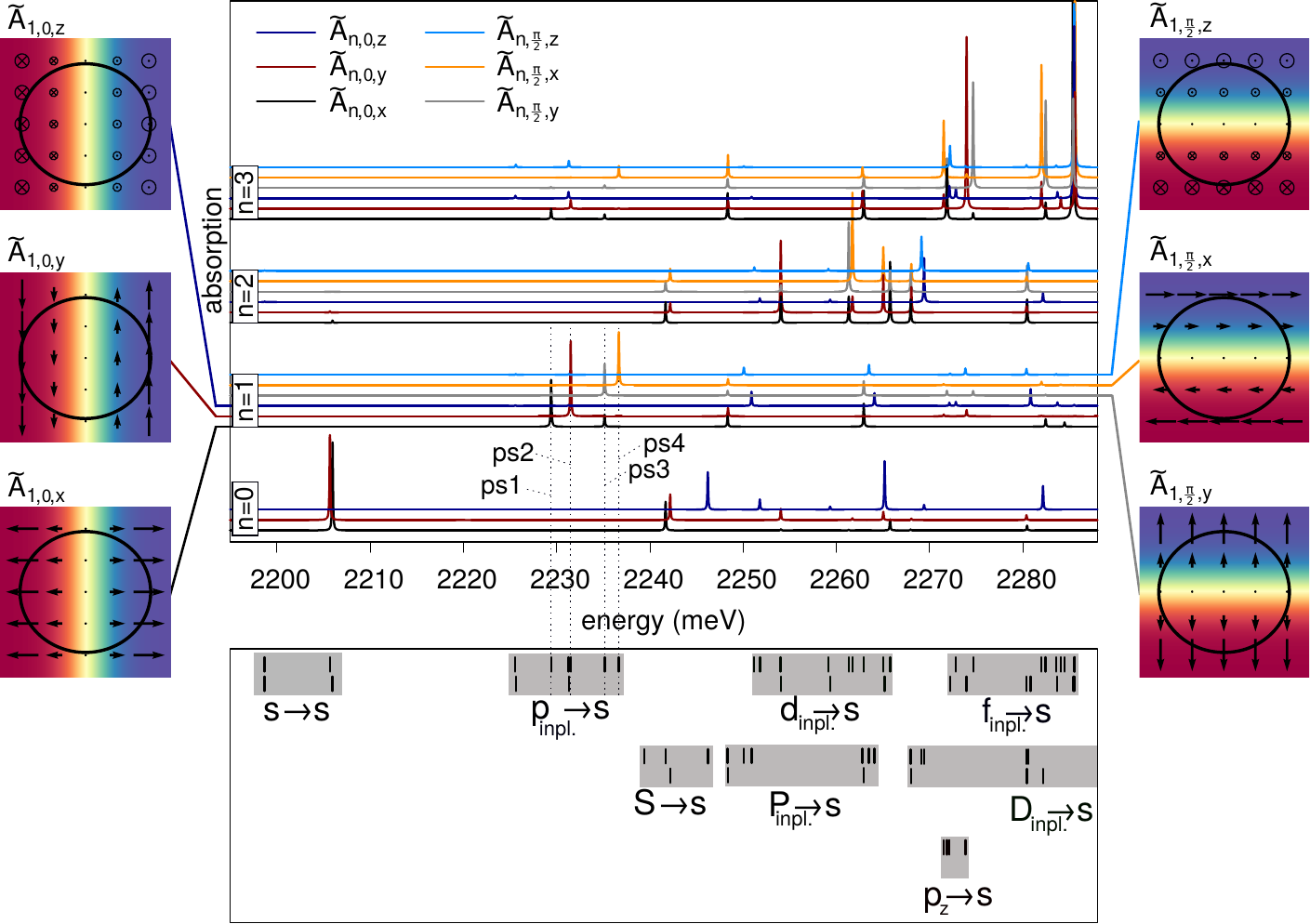}
\caption{Absorption spectra for a QD in our full model for different linear polarizations $\alpha$ and different rotations $\theta$. Below, all existing exciton eigenstates are displayed in blocks labeled by the most appropriate envelope basis state.}\label{fig:compare}
\end{figure}

\end{widetext}

Now we consider the full QD model by including Coulomb interaction and VBM. Furthermore we set $\beta=1.15$. The resulting level structure is given by dashes in the lower part of Fig.~\ref{fig:compare}.
In addition to a strong red-shift (the exciton binding energy, mainly caused by DCI), the previously four-fold spin degeneracy is completely lifted.
HH excitons like $s\to s$ are typically separated into two energetically close so called dark excitons (mainly consisting of $\epsilon_{{z}}$ and $\epsilon_{{{0}}}$) and two energetically close bright excitons (mainly consisting of $\epsilon_{y}$ and $\epsilon_{x}$) at higher energy.
LH excitons like $S\to s$ are typically separated into one single exciton (mainly consisting of $\epsilon_{\tilde{0}}$), two energetically close excitons (mainly consisting of $\epsilon_{\tilde{x}}$ and $\epsilon_{\tilde{y}}$) and a second single exciton at higher energy (mainly consisting of $\epsilon_{\tilde{z}}$).

Besides, the eigenstates loose their simple geometrical character. In other words, the considered interactions mix the different basis states so that an exciton eigenstate is never a pure e.g. $(s\to s)\epsilon_{x}$ state, but rather a mixture between for example $(s\to s)\epsilon_{x}$, $(d\to s)\epsilon_{x}$, $(S\to s)\epsilon_{\tilde{x}}$, ... basis states. Typically, one of the basis state contributions dominates and can be used to describe the main character of the eigenstate. In the following, we take this dominant contribution to label and refer to the eigenstates (in particular in Fig.~\ref{fig:compare}).

Absorption spectra are plotted in the main part of Fig.~\ref{fig:compare} for different orders of the beams up to $n=3$, different rotations $\theta$ and different polarizations $\vec{e}_\alpha$. In contrast to the reduced model (Fig.~\ref{fig:RedMod}), the absorption spectra for $x$-, $y$- and $z$-polarization are completely different, thus plotted separately. The light fields are drawn exemplarily on the left side of the spectra for the modes $\tilde{\vec{A}}_{1,0,x/y/z}$ and on the right side for the modes $\tilde{\vec{A}}_{1,\frac{\pi}{2},y/x/z}$.

The spectra are typically dominated by the peaks already observed in the reduced model.
Especially the $\text{HH}_{\text{inpl.}}\to s$ transitions can still be well identified.
Besides, some additional small lines appear due to the coupling of different basis states.
Although their oscillator strengths are typically small, in fact \textit{all} previously optically inaccessible eigenstates (envelope states with $\Delta a^z \neq 0$ as well as the dark HH and LH spin states) become optically accessible within the full model and for excitation with the appropriate light field.
To understand the details of the spectra, a better understanding of the state mixtures is necessary. Which states get mixed, is best answered by symmetry considerations:\\
With the reflections $\hat{R}_{\alpha}$ at a plane through the QD center with a normal in direction $\alpha \in \{x,y,z\}$ we can define the parity $P^{\alpha}$ by $\hat{R}_{\alpha} |\Psi \rangle = P^{\alpha} |\Psi \rangle$. Our model preserves a threefold reflection symmetry, accordingly basis states with different $P^{x}$, $P^{y}$ or $P^{z}$ are not coupled and our problem can be separated into eight subspaces, as they are listed in Tab.~\ref{tab:subspaces}. Thus, each eigenstate is a mixture of states from just one of the subspaces. Here we need to consider the symmetry of the full states, i.e., envelope and Bloch state combined. The parity of the Bloch states is given in Tab.~\ref{polarselection}, while the parity of the envelope states is given for exemplary states in Tab.~\ref{envselection}. In the notation of group theory, our model has $D_{2h}$-symmetry and the eigenstates are grouped according to the eight irreducible representations, as listed in Tab.~\ref{tab:subspaces}.

Each of the eight subspaces contains basis states which can be excited by appropriate light fields (see encircled states in Tab.~\ref{tab:subspaces}). Because all states in a subspace are mixed, all eigenstates become at least slightly optically active. Eigenstates addressable by in-plane polarized light (left column in Tab.~\ref{tab:subspaces}) are excitable by \textit{two different} light field geometries, what is studied in detail in Sec.~\ref{sec:dipolepolar}. A detailed discussion of the question which selection rules are broken by which interaction is given in App.~\ref{sec:appselecbreak}.

\begin{table}[h]
\centering
\begin{tabular}{c|c|c|}
\hline
\multirow{14}{*}{\centering \rotatebox{90}{\centering $n$ even}}&\tiny{$B_{2u}$: (E,O,E)} & \tiny{$A_{u}$: (O,O,O)} \\
&\fcolorbox{nix}{nix}{\textcolor{gray}{ (e,o,o) }$\textcolor{gray}{ \epsilon_{{z}}}        $} $\quad$ \fcolorbox{nix}{nix}{\textcolor{gray}{ (e,o,o) }$\textcolor{black}{\epsilon_{\tilde{z}}        }$} & \fcolorbox{nix}{nix}{\textcolor{black}{(o,o,e) }$\textcolor{gray}{ \epsilon_{{z}}}        $} $\quad$ \fcolorbox{ezo}{nix}{\textcolor{black}{(o,o,e) }$\textcolor{black}{\epsilon_{\tilde{z}}        }$} \\
&\fcolorbox{nix}{nix}{\textcolor{gray}{ (o,e,o) }$\textcolor{gray}{ \epsilon_{{{0}}}}$} $\quad$ \fcolorbox{nix}{nix}{\textcolor{gray}{ (o,e,o) }$\textcolor{gray}{ \epsilon_{\tilde{0}}}$} & \fcolorbox{nix}{nix}{\textcolor{black}{(e,e,e) }$\textcolor{gray}{ \epsilon_{{{0}}}}$} $\quad$ \fcolorbox{nix}{nix}{\textcolor{black}{(e,e,e) }$\textcolor{gray}{ \epsilon_{\tilde{0}}}$} \\
&\fcolorbox{exo}{nix}{\textcolor{black}{(o,o,e) }$\textcolor{black}{\epsilon_{x}        }$} $\quad$ \fcolorbox{exo}{nix}{\textcolor{black}{(o,o,e) }$\textcolor{black}{\epsilon_{\tilde{x}}}$} & \fcolorbox{nix}{nix}{\textcolor{gray}{ (e,o,o) }$\textcolor{black}{\epsilon_{x}        }$} $\quad$ \fcolorbox{nix}{nix}{\textcolor{gray}{ (e,o,o) }$\textcolor{black}{\epsilon_{\tilde{x}}}$} \\
&{\setlength{\fboxrule}{2pt} \fcolorbox{eye}{nix}{\textcolor{black}{(e,e,e) }$\textcolor{black}{\epsilon_{y}        }$}} $\,\,\,$ {\setlength{\fboxrule}{2pt} \fcolorbox{eye}{nix}{\textcolor{black}{(e,e,e) }$\textcolor{black}{\epsilon_{\tilde{y}}}$}} $\,$ & \fcolorbox{nix}{nix}{\textcolor{gray}{ (o,e,o) }$\textcolor{black}{\epsilon_{y}        }$} $\quad$ \fcolorbox{nix}{nix}{\textcolor{gray}{ (o,e,o) }$\textcolor{black}{\epsilon_{\tilde{y}}}$} \\
\cline{2-3}
&\tiny{$B_{3u}$: (O,E,E)} & \tiny{$B_{1u}$: (E,E,O)} \\
&\fcolorbox{nix}{nix}{\textcolor{gray}{ (o,e,o) }$\textcolor{gray}{ \epsilon_{{z}}}        $} $\quad$ \fcolorbox{nix}{nix}{\textcolor{gray}{ (o,e,o) }$\textcolor{black}{\epsilon_{\tilde{z}}        }$} & \fcolorbox{nix}{nix}{\textcolor{black}{(e,e,e) }$\textcolor{gray}{ \epsilon_{{z}}}        $} $\,\,\,\,$ {\setlength{\fboxrule}{2pt} \fcolorbox{eze}{nix}{\textcolor{black}{(e,e,e) }$\textcolor{black}{\epsilon_{\tilde{z}}        }$}}  \\
&\fcolorbox{nix}{nix}{\textcolor{gray}{ (e,o,o) }$\textcolor{gray}{ \epsilon_{{{0}}}}$} $\quad$ \fcolorbox{nix}{nix}{\textcolor{gray}{ (e,o,o) }$\textcolor{gray}{ \epsilon_{\tilde{0}}}$} & \fcolorbox{nix}{nix}{\textcolor{black}{(o,o,e) }$\textcolor{gray}{ \epsilon_{{{0}}}}$} $\quad$ \fcolorbox{nix}{nix}{\textcolor{black}{(o,o,e) }$\textcolor{gray}{ \epsilon_{\tilde{0}}}$} \\
&{\setlength{\fboxrule}{2pt} \fcolorbox{exe}{nix}{\textcolor{black}{(e,e,e) }$\textcolor{black}{\epsilon_{x}        }$}} $\,\,\,$ {\setlength{\fboxrule}{2pt} \fcolorbox{exe}{nix}{\textcolor{black}{(e,e,e) }$\textcolor{black}{\epsilon_{\tilde{x}}}$}} $\,$ & \fcolorbox{nix}{nix}{\textcolor{gray}{ (o,e,o) }$\textcolor{black}{\epsilon_{x}        }$} $\quad$ \fcolorbox{nix}{nix}{\textcolor{gray}{ (o,e,o) }$\textcolor{black}{\epsilon_{\tilde{x}}}$} \\
&\fcolorbox{eyo}{nix}{\textcolor{black}{(o,o,e) }$\textcolor{black}{\epsilon_{y}        }$} $\quad$ \fcolorbox{eyo}{nix}{\textcolor{black}{(o,o,e) }$\textcolor{black}{\epsilon_{\tilde{y}}}$} & \fcolorbox{nix}{nix}{\textcolor{gray}{ (e,o,o) }$\textcolor{black}{\epsilon_{y}        }$} $\quad$ \fcolorbox{nix}{nix}{\textcolor{gray}{ (e,o,o) }$\textcolor{black}{\epsilon_{\tilde{y}}}$} \\
\hline
\hline
\multirow{14}{*}{\centering \rotatebox{90}{\centering $n$ odd}}&\tiny{$A_{g}$: (E,E,E)} & \tiny{$B_{2g}$: (O,E,O)} \\
&\fcolorbox{nix}{nix}{\textcolor{gray}{ (e,e,o) }$\textcolor{gray}{ \epsilon_{{z}}}        $} $\quad$ \fcolorbox{nix}{nix}{\textcolor{gray}{ (e,e,o) }$\textcolor{black}{\epsilon_{\tilde{z}}        }$} & \fcolorbox{nix}{nix}{\textcolor{black}{(o,e,e) }$\textcolor{gray}{ \epsilon_{{z}}}        $} $\quad$ \fcolorbox{oze}{nix}{\textcolor{black}{(o,e,e) }$\textcolor{black}{\epsilon_{\tilde{z}}        }$} \\
&\fcolorbox{nix}{nix}{\textcolor{gray}{ (o,o,o) }$\textcolor{gray}{ \epsilon_{{{0}}}}$} $\quad$ \fcolorbox{nix}{nix}{\textcolor{gray}{ (o,o,o) }$\textcolor{gray}{ \epsilon_{\tilde{0}}}$} & \fcolorbox{nix}{nix}{\textcolor{black}{(e,o,e) }$\textcolor{gray}{ \epsilon_{{{0}}}}$} $\quad$ \fcolorbox{nix}{nix}{\textcolor{black}{(e,o,e) }$\textcolor{gray}{ \epsilon_{\tilde{0}}}$} \\
&\fcolorbox{oxe}{nix}{\textcolor{black}{(o,e,e) }$\textcolor{black}{\epsilon_{x}        }$} $\quad$ \fcolorbox{oxe}{nix}{\textcolor{black}{(o,e,e) }$\textcolor{black}{\epsilon_{\tilde{x}}}$} & \fcolorbox{nix}{nix}{\textcolor{gray}{ (e,e,o) }$\textcolor{black}{\epsilon_{x}        }$} $\quad$ \fcolorbox{nix}{nix}{\textcolor{gray}{ (e,e,o) }$\textcolor{black}{\epsilon_{\tilde{x}}}$} \\
&\fcolorbox{oyo}{nix}{\textcolor{black}{(e,o,e) }$\textcolor{black}{\epsilon_{y}        }$} $\quad$ \fcolorbox{oyo}{nix}{\textcolor{black}{(e,o,e) }$\textcolor{black}{\epsilon_{\tilde{y}}}$} & \fcolorbox{nix}{nix}{\textcolor{gray}{ (o,o,o) }$\textcolor{black}{\epsilon_{y}        }$} $\quad$ \fcolorbox{nix}{nix}{\textcolor{gray}{ (o,o,o) }$\textcolor{black}{\epsilon_{\tilde{y}}}$} \\
\cline{2-3}
&\tiny{$B_{1g}$: (O,O,E)} & \tiny{$B_{3g}$: (E,O,O)} \\
&\fcolorbox{nix}{nix}{\textcolor{gray}{ (o,o,o) }$\textcolor{gray}{ \epsilon_{{z}}}        $} $\quad$ \fcolorbox{nix}{nix}{\textcolor{gray}{ (o,o,o) }$\textcolor{black}{\epsilon_{\tilde{z}}        }$} & \fcolorbox{nix}{nix}{\textcolor{black}{(e,o,e) }$\textcolor{gray}{ \epsilon_{{z}}}        $} $\quad$ \fcolorbox{ozo}{nix}{\textcolor{black}{(e,o,e) }$\textcolor{black}{\epsilon_{\tilde{z}}        }$} \\
&\fcolorbox{nix}{nix}{\textcolor{gray}{ (e,e,o) }$\textcolor{gray}{ \epsilon_{{{0}}}}$} $\quad$ \fcolorbox{nix}{nix}{\textcolor{gray}{ (e,e,o) }$\textcolor{gray}{ \epsilon_{\tilde{0}}}$} & \fcolorbox{nix}{nix}{\textcolor{black}{(o,e,e) }$\textcolor{gray}{ \epsilon_{{{0}}}}$} $\quad$ \fcolorbox{nix}{nix}{\textcolor{black}{(o,e,e) }$\textcolor{gray}{ \epsilon_{\tilde{0}}}$} \\
&\fcolorbox{oxo}{nix}{\textcolor{black}{(e,o,e) }$\textcolor{black}{\epsilon_{x}        }$} $\quad$ \fcolorbox{oxo}{nix}{\textcolor{black}{(e,o,e) }$\textcolor{black}{\epsilon_{\tilde{x}}}$} & \fcolorbox{nix}{nix}{\textcolor{gray}{ (o,o,o) }$\textcolor{black}{\epsilon_{x}        }$} $\quad$ \fcolorbox{nix}{nix}{\textcolor{gray}{ (o,o,o) }$\textcolor{black}{\epsilon_{\tilde{x}}}$} \\
&\fcolorbox{oye}{nix}{\textcolor{black}{(o,e,e) }$\textcolor{black}{\epsilon_{y}        }$} $\quad$ \fcolorbox{oye}{nix}{\textcolor{black}{(o,e,e) }$\textcolor{black}{\epsilon_{\tilde{y}}}$} & \fcolorbox{nix}{nix}{\textcolor{gray}{ (e,e,o) }$\textcolor{black}{\epsilon_{y}        }$} $\quad$ \fcolorbox{nix}{nix}{\textcolor{gray}{ (e,e,o) }$\textcolor{black}{\epsilon_{\tilde{y}}}$} \\
\hline
\end{tabular}
\caption{Eight subspaces of exciton eigenstates defined by the three reflection symmetries in $x$-, $y$- and $z$-direction (corresponding to the eight irreducible representations of the $D_{2h}$-symmetry group). For each subspace, the parity in $x$-, $y$- and $z$-direction is given in capital letters (or the corresponding Mulliken symbol), followed by the eight corresponding combinations of the parities of envelope basis states (parities labeled in small letters) and parities of spin basis states (labeled by $\epsilon_{{z}}$, $\epsilon_{\tilde{z}}$ ...).
Optically accessible basis states are encircled following the color code:
$\textcolor{exe}{\tilde{\vec{A}}_{n,0,x}}$/
$\textcolor{exo}{\tilde{\vec{A}}_{n,\frac{\pi}{2},x}}$/
$\textcolor{eye}{\tilde{\vec{A}}_{n,0,y}}$/
$\textcolor{eyo}{\tilde{\vec{A}}_{n,\frac{\pi}{2},y}}$/
$\textcolor{eze}{\tilde{\vec{A}}_{n,0,z}}$/
$\textcolor{ezo}{\tilde{\vec{A}}_{n,\frac{\pi}{2},z}}$. The required parity of $n$ is given on the left. States accessible by plane wave like light are encircled by broader lines.
}
\label{tab:subspaces}
\end{table}

With the detailled listing of the state mixtures given in Tab.~\ref{tab:subspaces}, we now discuss the details of the absorption spectra in Fig.~\ref{fig:compare}. For this, we will focus on the states, which have been previously dark in the simplified model (cf. Fig.~\ref{fig:RedMod}).

The dark $(d_{yy}\to s)\epsilon_{{z}}$ exciton becomes bright due to its exceptionally strong mixture with the bright $(S\to s)\epsilon_{\tilde{z}}$ exciton. Accordingly, these two eigenstates dominate the spectrum for $n=0$ and $\vec{e}_z$-polarization (see blue spectrum for $n=0$ in Fig.~\ref{fig:compare}). This coupling can be utilized for example for an efficient excitation scheme of the dark exciton ground state \cite{holtkemper2020dark}.
More generally, all previously dark excitons $\sim \epsilon_{{{z}}/{{0}}/\tilde{0}}$ with even envelope parity in $z$-direction (subspaces $A_u$,$B_{1u}$,$B_{2g}$,$B_{3g}$) become slightly optically accessible by a coupling via VBM to the bright LH excitons $\sim \epsilon_{\tilde{z}}$. For each envelope, the spin directions $\sim \epsilon_{{{0}}/\tilde{0}}$ and $\sim \epsilon_{{{z}}/\tilde{z}}$ are separately accessible by the light fields $\tilde{\vec{A}}_{n,\frac{\pi}{2},z}$ and $\tilde{\vec{A}}_{n,0,z}$. The concrete assignment can be deduced from Tab.~\ref{tab:subspaces}. As an example, consider the $(d_{xx}\to s)\epsilon_{{{0}}}$ and $(d_{xx}\to s)\epsilon_{{z}}$ excitons, which are the energetically lowest two $d\to s$ states. $(d_{xx}\to s)\epsilon_{{{0}}}$, which has the symmetry $(e,e,e)\epsilon_{{{0}}}$ and is thereby related to $A_u$, is coupled to other states related to $A_u$, for instance to the bright states with symmetry $(o,o,e)\epsilon_{\tilde{z}}$, like $(D_{xy}\to s)\epsilon_{\tilde{z}}$. Similarly $(d_{xx}\to s)\epsilon_{{z}}$ is related to $B_{1u}$ and coupled to other states which are related to $B_{1u}$, for instance to the bright states $(S\to s)\epsilon_{\tilde{z}}$, $(D_{xx}\to s)\epsilon_{\tilde{z}}$ and $(D_{yy}\to s)\epsilon_{\tilde{z}}$. Accordingly, the energetically lower $(d_{xx}\to s)\epsilon_{{{0}}}$ is allowed in the spectrum for $\tilde{\vec{A}}_{2,\frac{\pi}{2},z}$. $(d_{xx}\to s)\epsilon_{{z}}$ is visible in both spectra $\tilde{\vec{A}}_{0,0,z}$ and $\tilde{\vec{A}}_{2,0,z}$. We note that the oscillator strength depends on the individual state and QD geometry \cite{holtkemper2018influence}, thus some excitons are not visible on a linear scale as used in Fig.~\ref{fig:compare}. As an example, the $(d_{yy}\to s)\epsilon_{{{z}}/{{0}}}$ in $\tilde{\vec{A}}_{2,0/\frac{\pi}{2},z}$ are very weak. Although the different spin states of one envelope are selectively addressable, different envelopes within one group are not separately accessible, e.g. $(d_{xx}\to s)\epsilon_{{z}}$, $(d_{xy}\to s)\epsilon_{{{0}}}$ and $(d_{yy}\to s)\epsilon_{{z}}$ are all addressable by $\tilde{\vec{A}}_{2,0,z}$.

\begin{widetext}

\begin{table}[H]
\centering
\begin{tabular}{c||c|c|c|c|c|c|c|c|c}
state & $s\to s$& $p_x\to s$&$p_y\to s$& $d_{xx}\to s$& $d_{xy}\to s$& $D_{xy}\to s$& $d_{yy}\to s$& $s\to p_x$ & $p_z\to s$ \\
\hline
 & & & & & & & & \\[-7pt]
parity $(P^x,P^y,P^z)$ & $(e,e,e)$ & $(o,e,e)$ & $(e,o,e)$ & $(e,e,e)$ & $(o,o,e)$ & $(o,o,e)$ & $(e,e,e)$ & $(o,e,e)$ & $(e,e,o)$ \\
\end{tabular}
\caption{Parity of the envelope part of exemplary electron-hole pair states. $o$ and $e$ refer to odd and even parity, respectively.}
\label{envselection}
\end{table}

\end{widetext}

Excitons $\sim \epsilon_{x/\tilde{x}}$ and $\sim \epsilon_{y/\tilde{y}}$ with odd $a^z$ (in the considered energetic range just the $(p_z\to s)\epsilon_{x/y}$, belonging also to the subspaces $A_u$,$B_{1u}$,$B_{2g}$,$B_{3g}$) get separately accessible by couplings to bright LH excitons $\sim \epsilon_{\tilde{z}}$ via the light fields $\tilde{\vec{A}}_{n,0,z}$ and $\tilde{\vec{A}}_{n,\frac{\pi}{2},z}$. Again, the concrete assignment can be deduced from the above symmetry considerations, as given in Tab.~\ref{tab:subspaces}.
The $(p_z\to s)\epsilon_{x}$ and $(p_z\to s)\epsilon_{y}$ excitons get bright in the case of $n=1$ by couplings to $(P_x\to s)\epsilon_{\tilde{z}}$ and $(P_y\to s)\epsilon_{\tilde{z}}$, respectively, which is visible in the bunch of small peaks for $n=1$.

Excitons $\sim \epsilon_{{{z}}/\tilde{z}}$ and $\sim \epsilon_{{{0}}/\tilde{0}}$ with odd $a^z$ (in the considered energetic range just the $(p_z\to s)\epsilon_{{{z}}/{{0}}}$, belonging to the subspaces $B_{2u}$,$B_{3u}$,$A_g$,$B_{1g}$) couple via VBM to bright LH excitons $\sim \epsilon_{\tilde{x}/\tilde{y}}$ and become optically accessible by in-plane polarized light. This is visible in the bunch of small peaks for $n=1$ at the appropriate energies. For a selective excitation see Sec.~\ref{sec:dipolepolar}.

Excitons $\sim \epsilon_{x/\tilde{x}}$ and $\sim \epsilon_{y/\tilde{y}}$ with even $a^z$ (belonging to the subspaces $B_{2u}$,$B_{3u}$,$A_g$,$B_{1g}$) are directly accessible by appropriate light fields and have been bright already in the reduced model. However, they are now coupled to each other and, thus, they are not exclusively addressable by the directly attributed light field, but also by a perpendicularly polarized and by $\frac{\pi}{2}$ rotated light field. This is displayed in Tab.~\ref{tab:subspaces}, where all subspaces, and all corresponding eigenstates, on the left are addressable by $\tilde{\vec{A}}_{n,0,x}$ and $\tilde{\vec{A}}_{n,\frac{\pi}{2},y}$ or by $\tilde{\vec{A}}_{n,\frac{\pi}{2},x}$ and $\tilde{\vec{A}}_{n,0,y}$.
As an example, it is instructive to consider the four $(p_{\text{inpl.}}\to s)\epsilon_{x/y}$ transitions:
They can be identified in Fig.~\ref{fig:compare} as those four states with the strongest absorption peak in the spectra for $\tilde{\vec{A}}_{1,0,x}$, $\tilde{\vec{A}}_{1,0,y}$, $\tilde{\vec{A}}_{1,\frac{\pi}{2},y}$ and $\tilde{\vec{A}}_{1,\frac{\pi}{2},x}$. 
They are coupled regarding $(p_x\to s) \epsilon_{x} \leftrightarrow (p_y\to s) \epsilon_{y}$ and $(p_x\to s) \epsilon_{y} \leftrightarrow (p_y\to s) \epsilon_{x}$ by an interplay between SRE and VBM. This leads to approximate eigenstates (from low to higher energy) 
\begin{align}
\text{state ps1: } & C_1 (p_x\to s) \epsilon_{x} + \tilde{C}_1 (p_y\to s) \epsilon_{y} \notag\\
\text{state ps2: } & C_2 (p_x\to s) \epsilon_{y} - \tilde{C}_2 (p_y\to s) \epsilon_{x} \notag\\
\text{state ps3: } & C_3 (p_y\to s) \epsilon_{y} - \tilde{C}_3 (p_x\to s) \epsilon_{x} \notag\\
\text{state ps4: } & C_4 (p_y\to s) \epsilon_{x} + \tilde{C}_4 (p_x\to s) \epsilon_{y} \, . \label{eq:eigenstp}
\end{align}
For a strongly elongated QD we get $C_i \gg \tilde{C}_i$, while for $L_x=L_y$ we get $C_i = \tilde{C}_i$.
In our case (VBM through Luttinger Hamiltonian without strain), the coupling $(p_x\to s) \epsilon_{y} \leftrightarrow (p_y\to s) \epsilon_{x}$ is weaker than $(p_x\to s) \epsilon_{x} \leftrightarrow (p_y\to s) \epsilon_{y}$ and the higher optical activity of $p_x \to s$ compared to $p_y\to s$ leads to the situation that just the admixture of $(p_x\to s)\epsilon_{x}$ to $(p_y\to s)\epsilon_{y}$ in state ps3 is strongly noticeable by an additional peak in the $\tilde{\vec{A}}_{1,0,x}$ spectrum.
Other $(\text{HH}_{\text{inpl.}}\to s)\epsilon_{x/y}$ couple in a similar way (for $d_{\text{inpl.}}\to s$ see App.~\ref{sec:appds}).

It is important to keep in mind that one main assumption within our model is a threefold reflection symmetry, i.e., a $D_{2h}$-symmetry. If the reflection symmetry is broken, the subspaces in Tab.~\ref{tab:subspaces} mix. For a broken reflection symmetry in $z$-direction, those subspaces with different parity in $z$-direction mix, here $[*,*,E]$ and $[*,*,O]$ (or $B_{2u}\leftrightarrow B_{3g}$ / $B_{3u}\leftrightarrow B_{2g}$ / $A_{u}\leftrightarrow B_{1g}$ / $B_{1u}\leftrightarrow A_{g}$), leading to a reduction to 4 subspaces, i.e., $C_{2v}$ symmetry. This is similar for a broken reflection symmetry in $x$- and $y$-direction. Thus, if the reflection symmetry is broken in all directions, all subspaces mix and all selection rules are broken. A quantitative description of the effect of symmetry breaking on the absorption spectra goes beyond the scope of this paper. However, for a reasonably small breaking of reflection symmetries, just small changes to the considered spectra are expected, as discussed in App.~\ref{sec:appsymmetrybreaking}.

\section{Quantitative measurement of the excitonic eigenstates}\label{sec:dipolepolar}

As we have seen in the previous sections, it is possible to identify certain eigenstates, e.g. the $\text{HH}_{\text{inpl.}}\to s$ transitions, from the absorption spectra of different cylindrical multipole modes. In fact, this identification of the spatial character of an eigenstate can be expanded to a quantitative measurement of the wave functions of all eigenstates. Here we emphasize, that such a measurement would go beyond electron density measurements by accessing the wave function itself, thus also the spatial phase information of the state. It is clear that the required experimental conditions are rather challenging. However, before discussing difficulties concerning the experimental realization in Sec.~\ref{sec:realis}, we focus in this section on the basic theoretical idea behind the measurement. 

The basic idea is to find a light field $\tilde{\vec{A}}(\vec{r}) = \sum_{m=1}^{M} a_m \tilde{\vec{A}}_m(\vec{r})$ with $\sum_{m=1}^{M} |a_m|^2=1$ which maximizes the absorption intensity of a considered eigenstate approximately described by $|\Psi\rangle \approx \sum_{n=1}^{N} c_n |n\rangle$ with $\sum_{n=1}^{N} |c_n|^2=1$ and $|n\rangle$ being a suitable set of orthogonal basis states. The absorption intensity (see Eq.~\eqref{eq:LMWW})
\begin{align}
\label{eq:dipolepolar11jhjh}
I &\sim \left|\langle \Psi | \hat{\tilde{H}}_{\gamma}(\tilde{\vec{A}}) | 0\rangle \right|^2 \notag\\
  &= \left| \sum_{m=1}^{M} \sum_{n=1}^{N} a_m \underbrace{\langle n |\hat{\tilde{H}}_{\gamma}(\tilde{\vec{A}}_m)|0\rangle}_{\alpha_{m,n}} c^*_n \right|^2 \, .
\end{align}
can be maximized with respect to the constraint $\sum_{m=1}^{M} |a_m|^2=1$ with the result \footnote{Equivalently, one could understand $a_m$ and $\sum_{n=1}^{N} \alpha_{m,n} c^*_n$ as the coefficients of two complex vectors, whose scalar product is maximal when they are parallel.}
\begin{align}
\label{eq:dipolepolar11}
a_m &= \frac{\sum_{n=1}^{N} \alpha^*_{m,n} c_n}{\sqrt{\sum_{m=1}^{M} \left|\sum_{n=1}^{N} \alpha^*_{m,n} c_n \right|^2}}
\end{align}
with $\alpha_{m,n}$ defined in Eq.~\eqref{eq:dipolepolar11jhjh}.
Thus, without normalization, the coefficients of the light field $a_m$ with maximal absorption into a state defined by the coefficients $c_n$ can be derived by a simple matrix multiplication $a_m = \sum_{n=1}^{N} \alpha^*_{m,n} c_n$.

To measure an eigenfunction, we propose to vary the light field, i.e. $a_m$, until a maximum in the absorption is found. From these $a_m$ one can derive the eigenstate, i.e. the coefficients $c_n$, if Eq.~\eqref{eq:dipolepolar11} can be inverted. Therefore we need to include at least $M=N$ multipole modes within the measurement.
A similar measurement is known for polarization sensitive absorption measurements, where the \textit{complex} coefficients $c_x$ and $c_y$ of a spin state $c_x \epsilon_{x} + c_y \epsilon_{y}$ can be deduced from the different orientations of linear polarization and the degree between linear and circular polarization. In this sense, we present a generalized polarization measurement, which accesses in addition to the spin degree of freedom also the spatial degree of freedom.

\begin{figure}[h]
\includegraphics[width=1.0\columnwidth]{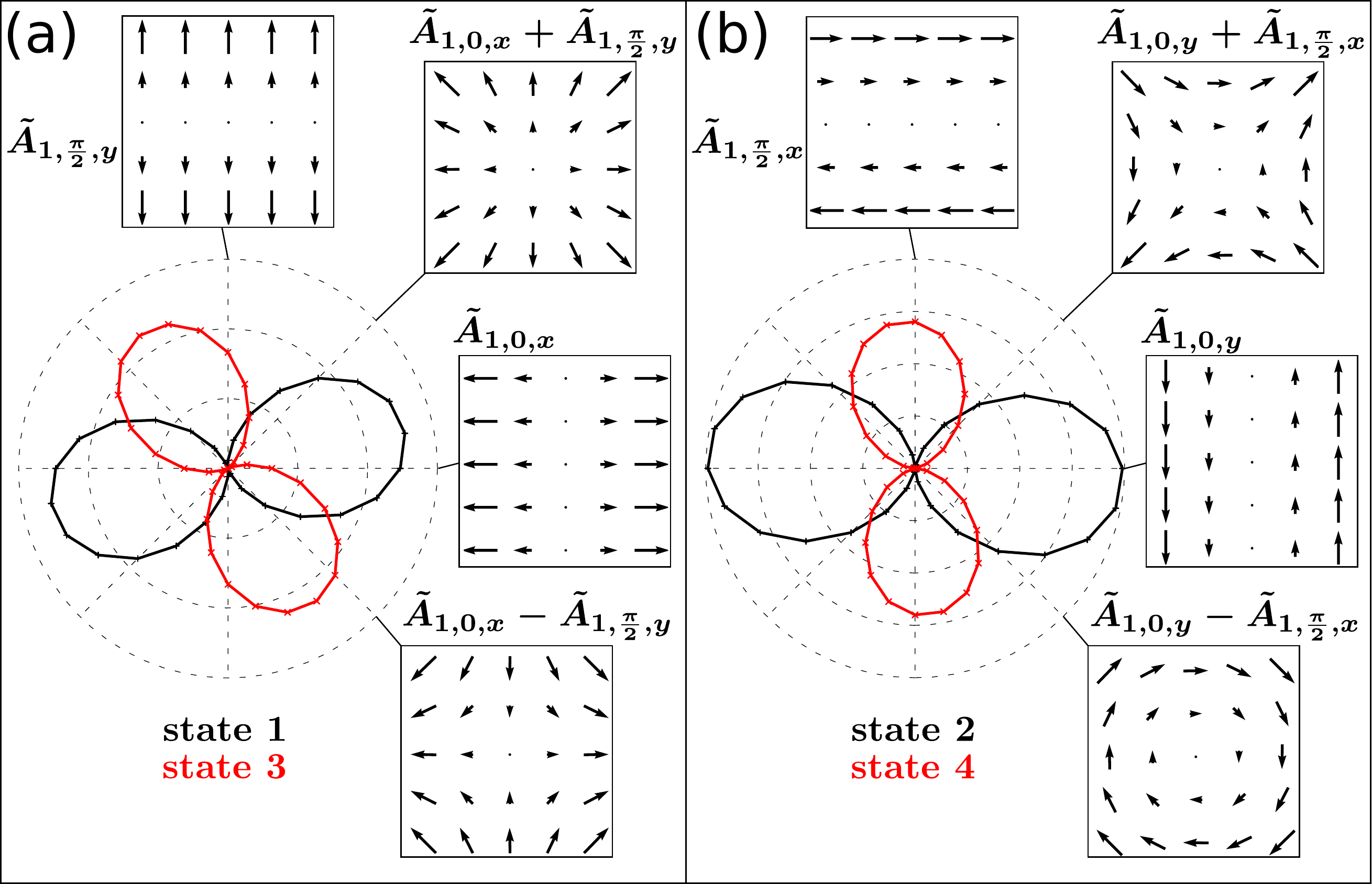}
\caption{Intensity of the $(p_{\text{inpl.}}\to s) \epsilon_{x/y}$ eigenstates for different superpositions of $\tilde{\vec{A}}_{1,0,x}$ and $\tilde{\vec{A}}_{1,\frac{\pi}{2},y}$ ($\tilde{\vec{A}}_{1,0,y}$ and $\tilde{\vec{A}}_{1,\frac{\pi}{2},x}$) in a (b). The insets display the corresponding field profiles.
}\label{fig:symmQDPolarps}
\end{figure}

The measurement is here demonstrated with the $(p_{\text{inpl.}}\to s)\epsilon_{x/y}$ excitons, while a second example is provided for $d_{\text{inpl.}}\to s$ in App.~\ref{sec:appds}. For a start we consider state ps3, which is the third state described in Def.~\eqref{eq:eigenstp} and, as seen in Fig.~\ref{fig:compare}, mainly excitable by the light fields $\tilde{\vec{A}}_{1,0,x}$ and $\tilde{\vec{A}}_{1,\frac{\pi}{2},y}$. Within our scheme, one would measure the absorption intensity for all possible superpositions of these two fields. However, we restrict our discussion to real-valued superpositions, because the maximal intensity is found for purely real superpositions, thus complex superpositions do not provide further insight and hinder visualization. The generalization to complex values is straight forward.
Real-valued superpositions to $a_1 \tilde{\vec{A}}_{1,0,x} + a_2 \tilde{\vec{A}}_{1,\frac{\pi}{2},y}$ include for example radially polarized fields and can be thought as continuous transitions between the fields plotted in the insets of Fig.~\ref{fig:symmQDPolarps} (a). In Fig.~\ref{fig:symmQDPolarps}, the absorption intensity of state ps3 is plotted in a polar plot with $\tan(\xi)=\frac{a_2}{a_1}$. As a result, the light field
\begin{align}
0.52\tilde{\vec{A}}_{1,0,x}-0.85\tilde{\vec{A}}_{1,\frac{\pi}{2},y} \notag
\end{align}
causes the strongest absorption of state ps3.
To obtain the related electronic state from this field, we need to assume an electronic basis. Since we just measured the absorption of two light modes, we need to restrict this basis to two states. Here we assume a basis built by the states $(p_x\to s) \epsilon_{x}$ and $(p_y\to s) \epsilon_{y}$, which are strongly excitable by the considered light fields. We could also assume higher basis states like $s\to p$, $P\to s$, $d\to p$ etc., which also couple noticeably to the measured light fields (see Fig.~\ref{fig:RedMod}). However, these basis states are at higher energies and not supposed to be strongly coupled to a state at the energy of ps3. If stronger contributions of these basis states would be assumed within the considered eigenstate, additional higher light modes need to be considered. $\alpha_{m,n}$ can be calculated from the basis. In our case, $\tilde{\vec{A}}_{1,0,x}$ just excites $(p_x\to s) \epsilon_{x}$ and $\tilde{\vec{A}}_{1,\frac{\pi}{2},y}$ just $(p_y\to s) \epsilon_{y}$ and we get \footnote{Different values for $\alpha_{11}$ and $\alpha_{22}$ arise, since we adjusted $p_x$ and $p_y$ to the QD elongation. We expect similar results and accuracy for non-adjusted basis states. Just the error caused by the contributions {$|\text{rest-bright}\rangle$} is assumed to rise slightly.} $\alpha_{\tilde{\vec{A}}_{1,0,x},(p_x\to s) \epsilon_{x}}=\alpha_{1,1}=1.0013$ and $\alpha_{\tilde{\vec{A}}_{1,\frac{\pi}{2},y},(p_y\to s) \epsilon_{y}}=\alpha_{2,2}=0.7441$.
From Eq.~\eqref{eq:dipolepolar11} we calculate $\frac{c_2}{c_1}=\frac{\alpha_{11}}{\alpha_{22}}\frac{a_2}{a_1}$. With this, the state deduced from the ``measurement'' would read
\begin{align}
|\text{ps3}_{\text{meas.}}\rangle &= 0.42 |(p_x\to s) \epsilon_{x}\rangle - 0.91|(p_y\to s) \epsilon_{y}\rangle \, . \notag
\end{align}
The ``exact'' state is
\begin{align}
\label{eq:exactstate3}
|\text{ps3}_{\text{exact}}\rangle =& \, 0.874 \Big[0.38 |(p_x\to s) \epsilon_{x}\rangle - 0.92|(p_y\to s) \epsilon_{y}\rangle\Big] \notag\\
 &+ 0.463|\text{rest-bright}\rangle + 0.149|\text{rest-dark}\rangle \, .
\end{align}
We find that the agreement between the ``measured'' and ``exact'' state is reasonably good. In particular the relative phase between the two basis states, i.e. the minus-sign, is correctly reproduced. Also the weights between the two states is well described by the measurement. The inaccuracy of the measurement is caused by contributions of higher basis states. $0.149^2 \approx 2\%$ are higher basis states which are optically not accessible ($|\text{rest-dark}\rangle$). Thus their coefficients cannot be accessed in general. $0.463^2 \approx 21\%$ are higher basis states which are optically accessible ($|\text{rest-bright}\rangle$). When considering higher light modes, most of their coefficients can be determined (however the coefficients of two basis states with interchanged particle states, like $p_x\to s$ and $s\to p_x$ for $\beta=1$, cannot be distinguished). Because some contributions of $|\text{rest-bright}\rangle$ are also addressable by the measured light modes, the relative contribution of $(p_x\to s) \epsilon_{x}$ and $(p_y\to s) \epsilon_{y}$ is slightly incorrect. The inaccuracy in terms of the angle $\xi$ between the ``measured'' eigenstate and the ``exact'' eigenstate in Eq.~\eqref{eq:exactstate3} without higher terms is less than 4\degree{}. This angle should not be misunderstood as a spatial direction, like in pure polarization measurements, but as a visualization of a superposition like on a Bloch sphere (in the complex case). In general, this Bloch sphere has $2M-1$ dimensions.

Similar measurements can be done for the other $(p_{\text{inpl.}}\to s)\epsilon_{x/y}$ states, as shown in Fig.~\ref{fig:symmQDPolarps}. The measured state ps1 would read
\begin{align}
|\text{ps1}_{\text{meas.}}\rangle &= 0.96 |(p_x\to s) \epsilon_{x}\rangle + 0.28|(p_y\to s) \epsilon_{y}\rangle \, . \notag
\end{align}
while the exact state reads
\begin{align}
|\text{ps1}_{\text{exact}}\rangle =& \, 0.839 \Big[0.95 |(p_x\to s) \epsilon_{x}\rangle + 0.32|(p_y\to s) \epsilon_{y}\rangle \Big] \notag\\
 &+ 0.527|\text{rest-bright}\rangle + 0.137|\text{rest-dark}\rangle \, .
\end{align}
Compared to state ps3, here we get a smaller and in-phase mixture of the basis states $(p_x\to s) \epsilon_{x}$ and $(p_y\to s) \epsilon_{y}$.\\
States ps2 and ps4 are nearly pure $(p_x\to s) \epsilon_{y}$ and $(p_y\to s) \epsilon_{x}$ basis states, respectively. From the measurement, one would get the states
\begin{align}
|\text{ps2}_{\text{meas.}}\rangle &= 1.00 |(p_x\to s) \epsilon_{y}\rangle - 0.09|(p_y\to s) \epsilon_{x}\rangle \notag\\
|\text{ps4}_{\text{meas.}}\rangle &= 0.07 |(p_x\to s) \epsilon_{y}\rangle - 1.00|(p_y\to s) \epsilon_{x}\rangle \, . \notag
\end{align}
while the ``exact'' states read
\begin{align}
|\text{ps2}_{\text{exact}}\rangle =& \, 0.886 \Big[1.00 |(p_x\to s) \epsilon_{y}\rangle - 0.01|(p_y\to s) \epsilon_{x}\rangle\Big] \notag\\
 &+ 0.447|\text{rest-bright}\rangle + 0.125|\text{rest-dark}\rangle \notag\\
|\text{ps4}_{\text{exact}}\rangle =& \, 0.886 \Big[0.01 |(p_x\to s) \epsilon_{y}\rangle + 1.00|(p_y\to s) \epsilon_{x}\rangle\Big] \notag\\
 &+ 0.442|\text{rest-bright}\rangle + 0.139|\text{rest-dark}\rangle \, .
\end{align}
For all $p_{\text{inpl.}}\to s$ we find that our proposal describes the exciton wave function between the participating states within resonable accuracy. We stress that our proposal can retrieve the relative phases between the different spatial contribution, i.e., the phase field or wave function of the exciton. Thereby, we go beyond measurements of electron densities.

When using our proposal, one should keep the following things in mind:
\begin{enumerate}
\item{A global phase is - of course - not accessible.}
\item{We are measuring the excitonic wave function, but it is not determinable whether the electron or the hole is in a certain state. In other words, we can not distinguish between contributions of e.g. $p_x\to s$ and $s\to p_x$ basis states for $\beta=1$. To understand this statement, we should recapitulate that the envelope part of the light-matter interaction is defined by $$\vec{M}_{a_1,a_2}(\tilde{\vec{A}})=\int d^3R\,\,\Phi_{a_1}^*(\vec{R})\,\Phi_{a_2}(\vec{R})\, \tilde{\vec{A}}(\vec{R})$$ (see Eq.~\eqref{eq:LMWW}). Thus, the light field does not ``see'' the electron-hole-pair envelope basis state $$\quad \quad \Phi^{\text{exciton}}_{a_1,a_2}\left(\vec{r}_{\text{hole}},\vec{r}_{\text{elec.}}\right) = \Phi^{\text{hole}*}_{a_1}\left(\vec{r}_{\text{hole}}\right) \Phi^{\text{elec.}}_{a_2}\left(\vec{r}_{\text{elec.}}\right)\, ,$$ but ``just'' the wave function $\Phi^{\text{exciton}}_{a_1,a_2}\left(\vec{r}_{\text{hole}},\vec{r}_{\text{elec.}}\right)$ with $\vec{r}_{\text{hole}}=\vec{r}_{\text{elec.}}$. For $\beta=1$ and real envelope functions, we get $\Phi^{\text{exciton}}_{a_1,a_2}\left(\vec{r},\vec{r}\right) = \Phi^{\text{exciton}}_{a_2,a_1}\left(\vec{r},\vec{r}\right)$. Thus it is not determinable whether the electron is in state $a_1$ and the hole in state $a_2$, or the other way round. The requirement $\beta = 1$ defines just a convenient basis, where it is obvious that $\alpha_{m,n}$ is never invertible for e.g. $p_x\to s$ and $s\to p_x$ basis states and any set of light fields. $\beta\neq 1$ does not change this statement in general, but just introduces non-orthonormal basis states for electrons and holes which make it harder to see whether $\alpha_{m,n}$ is invertible.}
\item{The coefficients of dark basis states are not determinable, also due to the required invertibility of $\alpha_{m,n}$.}
\item{The measurement is not restricted by the diffraction limit.}
\end{enumerate}

There are several possible applications closely related to this measurement:
\begin{enumerate}
\item{One can tune the light field geometry to get a maximal absorption into a certain state.}
\item{The measurement of the exciton wave function allows to draw conclusions about the QDs geometry.}
\item{A standard polarization sensitive absorption measurement allows the readout of the spin degree of freedom of an electronic state within a QD. Such a measurement is typically the basis to use the spin as a storage for quantum information. In the same sense, the proposed generalized polarization measurement could pave the way to use the infinite spatial degree of freedom to store quantum information within a QD. Here, the problem of quickly decaying higher exciton states requires further research. However, at least for $n=1$ relatively long lived metastable trion triplet states \cite{hinz2018charge} are available.}
\item{The selectivity of an optical excitation can be increased by a suitable superposition of cylindrical multipole modes. This is equivalent to polarization-selective excitation when energetic selectivity is not possible, and might be similarly powerful.}
\end{enumerate}

It might be instructive to consider this measurement from an alternative perspective: We can measure the absorption intensity (neglecting the spin)
\begin{align}
\label{eq:jhgkjhblbjh}
\Big| \int d^3R\underbrace{\Phi_{\text{elec.}}^*(\vec{R})\Phi_{\text{hole}}(\vec{R})}_{\Phi_{\text{exc.}}(\vec{R},\vec{R})} \tilde{\vec{A}}(\vec{R}) \Big|^2
\end{align}
for arbitrary light fields $\tilde{\vec{A}}(\vec{R})$. The maximal intensity is expected for $\tilde{\vec{A}}(\vec{R})=\Phi_{\text{exc.}}^*(\vec{R},\vec{R})$, thus the full complex exciton wave function $\Phi_{\text{exc.}}(\vec{R},\vec{R})$ can be measured by tuning the form of the light field until the intensity is maximized.

\section{Realistic light fields}\label{sec:realis}

\subsection{Notes on the experimental feasibility}\label{sec:reali}

To estimate the experimental feasibility of our proposal, we discuss the intensity of the absorption for higher $n$. Throughout this paper, the amplitude of the light fields is $\sim\left(\frac{r}{R_{\text{QD}}}\right)^n$ (Eq.~\eqref{eq:ideallightE}). $r$ leads to contributions to the light-matter coupling strength in the order of $R_{\text{QD}}$, resulting in a scaling of the intensity $\sim\left(\frac{R_{\text{QD}}}{R_{\text{QD}}}\right)^{2n}=1$, resulting in the comparable intensities of the different orders.
However, the natural scaling can be deduced, for instance, from Bessel beams and is proportional to $\frac{\left(q_r r\right)^{n}}{2^n n!}$ (see App.~\ref{sec:appBesToNod}). The expected intensity for a light field of order $n$ is thus $\sim \left(\frac{\left(q_r R_{\text{QD}}\right)^{n}}{2^n n!}\right)^2$.\\
$q_r$ is fixed via $\sqrt{q_r^2+q_z^2} = \frac{n_r E}{\hbar c_0}$ with the excitation energy $E = 2.2$~eV, a factor describing the beam width $\frac{q_r}{q_z} = 1$ \footnote{$\frac{q_r}{q_z} = 1$ is a realistic value, as shown in Refs.~\cite{chen2009realization,huang2015efficient}} and the refractive index of CdSe of $n_r \approx 2.8$ \cite{ninomiya1995optical} to $q_r\approx \frac{1}{45 \text{~nm}}$. With an average QD radius of $R_{\text{QD}} = 2.7$~nm and fixed $A_0$, the intensity of the peaks is reduced by a factor $\sim 10^{-3}$, $\sim 10^{-7}$, $\sim 10^{-11}$ ... with order $n=1,2,3,...$ of the light mode. We should highlight, that optimizations are possible in regard of $\frac{q_r}{q_z}$, the QD radius $R_{\text{QD}}$, the excitation energy $E$, the diffraction index $n_r$ or by turning up $A_0$ for higher $n$. Not just far field beams, but also near fields created e.g. by laser-illuminated metal tips \cite{zurita2002multipolare} could be used.

The proposed measurements of the wave function in Sec.~\ref{sec:dipolepolar} requires in general a combination of light field modes of different order. Since the oscillator strength of these light modes differs by several orders of magnitude, the proposed measurement requires in general a very accurate adjustment of the light mode intensities over several orders of magnitude. However, in the special case of the presented measurement in Sec.~\ref{sec:dipolepolar}, just light modes of the same order (here $n=1$) are utilized, thus ``just'' a good suppression of the $n=0$ modes is required.

The alignment between QD center and the center of the beam profile is important. If we shift the light field around $R_{\text{misfit}}$ away from the QD center, we get the substitution $(\frac{r}{R_{\text{QD}}})^n \to (\frac{r-R_{\text{misfit}}}{R_{\text{QD}}})^n$ which causes for example the additional plane wave like absorption peaks with a relative intensity in the order of $(\frac{R_{\text{misfit}}}{R_{\text{QD}}})^n$. Therefore for pure light modes of order $n$ with a misalignment of $R_{\text{misfit}}=R_{\text{QD}}$, we already expect similar absorption intensities of plane wave like absorption patterns and the intended patterns for the light field of order $n$.

\subsection{Solenoidal fields}\label{sec:ps}
Up to now we considered just convenient components of light fields. Realistic light fields need to be solenoidal (in free space and Coulomb gauge $\text{div}\left(\vec{A}\right)=0$ has to be fullfiled), what is clearly not the case for modes like $\tilde{\vec{A}}_{1,0,x}(\vec{r},t) \sim x \vec{e}_{x}$. A well defined theoretical basis to describe light fields is given by Bessel beams, which constitute an exact and complete solution of the vectorial Helmholtz equation. Although Bessel beams have an infinite radial extension, there are several experimentally realized approximations \cite{durnin1987diffractionfree,woerdemann2013advanced,ettorre2015experimental}. Around the beam axis, Bessel beams can be described by a simple superposition of few cylindrical multipole modes. In the following, we will discuss these superpositions for several different representations of Bessel beams, which are given explicitly in App.~\ref{sec:appBesToNod}.

First, we consider propagating beams, i.e., propagating light fields with a finite extension of their beam profile. For an exact representation, one always has to combine a component polarized perpendicular to the propagation direction, i.e. polarized in $x$/$y$-direction, and additionally a component polarized along the propagation direction, i.e., the $z$-direction). In our case, we need to combine the modes $\tilde{\vec{A}}_{n,\theta,x}$ with $\frac{q_r}{q_z}\tilde{\vec{A}}_{n\pm1,\theta,z}$ as well as the modes $\tilde{\vec{A}}_{n,0/\frac{\pi}{2},y}$ with $\frac{q_r}{q_z}\tilde{\vec{A}}_{n\pm 1,\frac{\pi}{2}/0,z}$ (see Eqs.~\eqref{eq:Bessel_x_y} in App.~\ref{sec:appBesToNod}).
While the modes $\tilde{\vec{A}}_{n,\theta,x/y}$ are scaled on the order of $(q_r R_{\text{QD}})^{n}$, the modes $\frac{q_r}{q_z}\tilde{\vec{A}}_{n\pm1,\theta,z}$ are on the order of $\frac{q_r}{q_z}(q_r R_{\text{QD}})^{n-1}$. Accordingly $\tilde{\vec{A}}_{n,\theta,x/y}$ is about a factor $q_z R_{\text{QD}}$ smaller than $\frac{q_r}{q_z}\tilde{\vec{A}}_{n\pm1,\theta,z}$. Because $q_z$ has an upper limit of $\frac{n_r E}{\hbar c_0}$ even for the undesired case $\frac{q_r}{q_z}=0$, we get $q_z R_{\text{QD}} < 0.08$. Therefore always the term polarized in $z$-direction dominates at the beam axis (except for $n=0$). For our standard case $\frac{q_r}{q_z}=1$ we get $q_z R_{\text{QD}} \approx 0.06$. The absorption spectra of the combined realistic modes can be deduced by a combination of the spectra for $\tilde{\vec{A}}_{n,\theta,x}$ and $\tilde{\vec{A}}_{n\pm1,\theta,z}$ ($\tilde{\vec{A}}_{n,0/\frac{\pi}{2},y}$ and $\tilde{\vec{A}}_{n\pm 1,\frac{\pi}{2}/0,z}$). If a threefold reflection symmetry is present, we can simply add the spectra, while with broken reflection symmetry in $z$-direction, an eigenstate might be addressable by both, $x$/$y$- and $z$-polarized modes, where constructive/destructive interference needs to be considered. This combination of spectra reduces the possibility for selective excitation. States excitable by $\tilde{\vec{A}}_{n,\theta,x}$ and $\tilde{\vec{A}}_{n\pm1,\theta,z}$ ($\tilde{\vec{A}}_{n,0/\frac{\pi}{2},y}$ and $\tilde{\vec{A}}_{n\pm 1,\frac{\pi}{2}/0,z}$) cannot be accessed separately anymore; for example consider state ps1 (addressable by $\tilde{\vec{A}}_{1,0,x}$) and the $(S\to s)\epsilon_{\tilde{z}}$-like eigenstate (addressable by $\tilde{\vec{A}}_{0,0,z}$). Furthermore, the states addressable by the much stronger $z$-polarized fields will always dominate the spectra. The measurement of the wave functions via Eq.~\eqref{eq:dipolepolar11} is not affected if there is a reflection symmetry in $z$-direction. In that case, there is no coupling between eigenspaces excitable by light polarized in $x$/$y$-direction and $z$-direction (see Tab.~\ref{tab:subspaces}). If the reflection symmetry in $z$-direction is broken, such couplings are possible. For example, a small admixture of $(S\to s)\epsilon_{\tilde{z}}$ (excitable by $\tilde{\vec{A}}_{0,0,z}$) occurs within state ps1 (excitable by $\tilde{\vec{A}}_{1,0,x}$ and $\tilde{\vec{A}}_{1,\frac{\pi}{2},y}$). As we saw, the mode $\tilde{\vec{A}}_{1,0,x}$ is always superposed with a much stronger mode $\tilde{\vec{A}}_{0,0,z}$, thus the small admixture of $(S\to s)\epsilon_{\tilde{z}}$ can have a significant impact on the absorption intensity, what would distort the measurement as presented in Sec.~\ref{sec:dipolepolar}. To restore a correct measurement of the wave function, one could include a third light field mode (e.g. the combined field of $\tilde{\vec{A}}_{3,\frac{\pi}{2},y}$ and $\tilde{\vec{A}}_{2,0,z}$, which mainly excites $(S\to s)\epsilon_{\tilde{z}}$) into the measurement and explicitly consider the $(S\to s)\epsilon_{\tilde{z}}$ basis state in the evaluation of Eq.~\eqref{eq:dipolepolar11}. We conclude that the measurement is not hindered in general, but requires more effort.

The omnipresent strong $z$-polarized component vanishes in certain cases within another representation of Bessel beams. To describe such a representation, we first define an alternative multipole expansion via

{\footnotesize
\begin{align}
& \tilde{\vec{A}}_{n,\theta,xy}(\vec{r},t)         = A_{0} \left( \frac{r}{R_\text{QD}} \right)^n \left[ \cos(n \varphi-\theta) \, \vec{e}_{x} - \sin(n \varphi-\theta) \, \vec{e}_{y} \right] \notag\\
& \tilde{\vec{A}}_{n,\theta,\tilde{xy}}(\vec{r},t) = A_{0} \left( \frac{r}{R_\text{QD}} \right)^n \left[ \cos(n \varphi-\theta) \, \vec{e}_{x} + \sin(n \varphi-\theta) \, \vec{e}_{y} \right] \, . \label{eq:ideallightExy}
\end{align}
}
These modes still have to be combined with $z$-polarized modes to obtain realistic fields. We have to combine the modes $\tilde{\vec{A}}_{n,\theta,xy}$ with $\tilde{\vec{A}}_{n+1,\theta,z}$ as well as the modes $\tilde{\vec{A}}_{n+1,\theta,y}$ with $\tilde{\vec{A}}_{n,\theta,z}$ (see Eqs.~\eqref{eq:Bessel_xy_z} in App.~\ref{sec:appBesToNod}). Thus, in the first case the terms with $x$/$y$-polarization $\tilde{\vec{A}}_{n,\theta,xy}$ dominate and the $z$-polarized component is negligible, while in the second case the $z$-polarized component still prevails (except for $n=0,\theta=\frac{\pi}{2}$). The above described problems with the undesired admixed modes is just banned for certain special cases.

To evade the undesired mixture of different multipole modes, one can consider standing waves. In fact, standing waves are created standardly in various micro-cavities built around QDs to increase the light matter coupling \cite{schneider2016quantum,reithmaier2004strong}. Higher Hermite-Gaussian and Laguerre-Gaussian modes were found in hemispherical micro-cavities \cite{cui2006hemispherical}. For standing waves we can position the QD into a nodal line of the field in $z$-direction and get the fields (see Eq.~\eqref{eq:Bessel_x_y_standing_approx} in App.~\ref{sec:appBesToNod}):

{\footnotesize
\begin{align}
& \tilde{\vec{\breve{A}}}^{x}_{n,\theta}(\vec{r},t) = A_{0}(t) \frac{r^n}{R_\text{\tiny QD}^n} \left( \cos(n \varphi-\theta) \, \vec{e}_{x} - \frac{n z}{r} \cos((n-1) \varphi-\theta) \, \vec{e}_{z} \right) \notag\\
& \tilde{\vec{\breve{A}}}^{y}_{n,\theta}(\vec{r},t) = A_{0}(t) \frac{r^n}{R_\text{\tiny QD}^n} \left( \cos(n \varphi-\theta) \, \vec{e}_{y} + \frac{n z}{r} \sin((n-1) \varphi-\theta) \, \vec{e}_{z} \right) \notag\\
& \tilde{\vec{\breve{A}}}^{z}_{n,\theta}(\vec{r},t) = A_{0}(t) \frac{r^n}{R_\text{\tiny QD}^n} \cos(n \varphi-\theta) \, \vec{e}_{z} \label{eq:ideallightEstand}
\end{align}
}
The additional terms $\sim z \vec{e}_z$ are of the same order as the $x$/$y$-polarized terms and induce transitions into LH excitons excited in $z$-direction, like $P_{z}\to s$, $D_{xz}\to s$/$D_{yz}\to s$, $F_{xxz}\to s$/$F_{xyz}\to s$/$F_{yyz}\to s$ etc. . Those states are well above the considered energetic range and weakly coupled to the studied transitions. Thus the absorption spectra of the previous sections hold in good agreement also for the realistic light fields of Eq.~\eqref{eq:ideallightEstand}.

\section{Conclusion}\label{sec:conclusion}
We have analyzed the absorption of spatially structured light beams by a QD.
We focussed on cylindrical multipole transitions and derived analytical selection rules for a simplified QD model. We have studied the coupling mechanisms via Coulomb interactions and valence band mixing, which lead to the optical addressability of all electronic eigenstates of the QD. Within this extended model, we analyzed the possibility to tailor the optical activity of certain states by varying the spatial shape of the exciting light field. Thereby we explored a way to excite the QDs eigenstates selectively, without the need of spectral separation. This way is similar to spin selective excitation and could help to overcome today's limits in time resolution of certain pump probe experiments. Furthermore, we proposed a method to measure the excitonic wave function of arbitrary eigenstates, including relative spatial phases and thereby going beyond electron density measurements. Such a measurement is the prerequisite to use the infinite spatial degree of freedom for QD based quantum information technology.
We explored the measurement of the wave function of the first excited exciton states in detail and estimated the precision. The experimental feasibility of the proposed techniques as well as different possibilities to realize cylindrical multipole modes are discussed by a comparison with different representations of Bessel beams.

\begin{acknowledgments}
G. F. Q. thanks the ONRG for financial support through NICOP grant N62909-18-1-2090.
\end{acknowledgments}

\appendix

\section{Similarity between the matrix elements of $\hat{\vec{p}}$ and $\hat{\vec{r}}$}\label{sec:appeqalpr}

Using the identity $\hat{p}_{\alpha}=-i \frac{m_0}{\hbar}[\hat{r}_{\alpha},\hat{H}_0]$ with $\hat{H}_0=\frac{\hat{\vec{p}}^2}{2m}+V(\hat{\vec{r}})$ being the Hamiltonian within the bulk crystal as well as the assumption that $\langle \hat{\vec{r}} |b\rangle = u_{b}(\vec{r})$ are eigenfunctions of $\hat{H}_0$ (at the $\Gamma$-point) and $E_b$ the corresponding eigenenergies, we deduce that
\begin{align*}
\mu^{\alpha}_{b_1,b_2} = e \langle b_1 |\frac{\hat{p}_{\alpha}}{m_0}| b_2 \rangle &= - i e \frac{1}{\hbar}\langle b_1 |[\hat{r}_{\alpha},\hat{H}_0]| b_2 \rangle \notag\\
&= - i e \frac{1}{\hbar}\langle b_1 |\hat{r}_{\alpha}\hat{H}_0 - \hat{H}_0\hat{r}_{\alpha}| b_2 \rangle \notag\\
&= - i e \frac{E_{b_2}-E_{b_1}}{\hbar} \langle b_1 |\hat{r}_{\alpha}| b_2 \rangle
\end{align*}
with $-e \langle b_1 |\hat{r}_{\alpha}| b_2 \rangle$ representing the dipole moments in the typical form.

\section{Analytical selection rules and limits of the selectivity}\label{sec:appderanaselect}

The optical selection rules between envelopes described by Hermite-Gaussian functions $\Phi_{a}$ are requested. Therefore we need to solve the integral (see Eq.~\eqref{eq:LMWW})
\begin{align*}
\vec{M}_{a_1,a_2}(\tilde{\vec{A}}) = \int d^3R\,\,\Phi_{a_1}^*(\vec{R})\,\tilde{\vec{A}}(\vec{R})\, \Phi_{a_2}(\vec{R}) \, . \notag
\end{align*}
For certain expansions of the light field $\tilde{\vec{A}}(\vec{R})$, compact analytical selection rules can be given.

\underline{Power functions:} For an expansion into Power functions $\tilde{\vec{A}}^{\text{Power}}_{n_x,n_y,n_z}(\vec{R}) = \tilde{\vec{A}}_0 x^{n_x} y^{n_y} z^{n_z}$, the integral can be decomposed into three one-dimensional integrals, which can be solved by ladder operator algebra. With $\Delta a^{\alpha} = |a^{\alpha}_{\text{hole}}-a^{\alpha}_{\text{elec.}}|$, the selection rules read:
\begin{enumerate}
\item[1.]{The parity of $\Delta a^{\alpha}$ and $n_{\alpha}$ has to be the same.}
\item[2.]{$\Delta a^{\alpha} \leq n_{\alpha}$}
\end{enumerate}
It is possible to tailor selection rules within certain limits. As an example, it is possible to deactivate an arbitrary transition $a^{\alpha} \to \tilde{a}^{\alpha}$ by an adequate superposition of the power functions $\alpha^{|a-\tilde{a}|}$ and $\alpha^{|a-\tilde{a}|+2}$.

\underline{Hermite Polynomials:} The question arises, whether it might be possible to increase the selectivity by an optimized set of light fields. To explore the theoretical limit of such a selectivity, it is instructive to expand the light fields into Hermite polynomials $H_n$ via $\tilde{\vec{A}}^{\text{Hermite}}_{n_x,n_y,n_z}(\vec{R}) = \tilde{\vec{A}}_0 H_{n_x}(\frac{2 x}{L_x}) H_{n_y}(\frac{2 y}{L_y}) H_{n_z}(\frac{2 z}{L_z})$. These functions are similar to the envelopes of the electronic basis states in the QD.
Again, we can decompose the integral into one-dimensional problems. In addition to the above mentioned selection rules, we get:
\begin{enumerate}
\item[3.]{$a^{\alpha}_{\text{elec.}} + a^{\alpha}_{\text{hole}} \geq n_{\alpha}$.}
\end{enumerate}
Therewith, for example the transitions between the envelopes $a_{\text{hole}}^{\alpha}=n_{\alpha}$ and $a_{\text{elec.}}^{\alpha}=0$ are just accessible by the light field $\sim H_{n_{\alpha}}(\frac{2 \alpha }{L_{\alpha}})$, respectively for each $n_{\alpha}$. Thus any superposition of Hermite polynomials would lead to less restrictive selection rules regarding those states. In particular, it is not possible to find light fields which just address one transition or increase the selectivity to $\Delta a^{\alpha} = n_{\alpha}$.\\
$\tilde{\vec{A}}^{\text{Hermite}}$ includes unrealistically small radial variations in the order of the QD size. Thus it is just discussed in this theoretical paragraph.

\underline{Cylindrical multipole modes:} For the cylindrical multipole modes discussed in the main part of the paper, the selection rules can be deduced from the above findings via an expansion into Hermite polynomials
{\small
\begin{align}
r^n \cos(n \varphi) =& \sum_{m=0}^{\frac{n}{2}} (-1)^m \binom{n}{2m} H_{n-2m}(x) H_{2m}(y) \label{eq:ideallightECos}\\
r^n \cos(n \varphi-\frac{\pi}{2}) =& \notag\\
 =\sum_{m=1}^{\frac{n+1}{2}} (-1&)^{m-1} \binom{n}{2m-1} H_{n-2m+1}(x) H_{2m-1}(y) . \notag
\end{align}
}
The resulting selection rules (see Sec.~\ref{sec:reducedmodelab}) are similar to those of single Hermite polynomials, but lack the independent validity in each direction.

\section{Which selection rules are broken by which interaction?}\label{sec:appselecbreak}
It might be instructive to understand for each interaction, which selection rule it breaks or preserves:\\

\underline{DCI:} DCI has no effect on spins and thus no effect on spin selection rules. However, it causes mixtures between envelopes: The two electron-hole pair basis states described by $a_{\text{elec.}}^{\alpha}$ and $a_{\text{hole}}^{\alpha}$ as well as $\tilde{a}_{\text{elec.}}^{\alpha}$ and $\tilde{a}_{\text{hole}}^{\alpha}$ get mixed, if $(a_{\text{elec.}}^{\alpha} + a_{\text{hole}}^{\alpha}) - (\tilde{a}_{\text{elec.}}^{\alpha} + \tilde{a}_{\text{hole}}^{\alpha}) \in \{0, 2, 4, ... \}$. This breaks the third envelope selection rule and reduces the second envelope selection rule to the statement, that ``the parity of $\Delta a^z$ has to be even'' (see Sec.~\ref{sec:reducedmodelab}). In consequence, for example weak $d_{xx}\to s$ transitions become allowed already for plane wave like light ($n=0$).

\underline{$\beta \neq 1$:} Different confinement lengths ($\beta \neq 1$) lead to non-orthogonal sets of envelope basis functions for electrons and holes. Therefore the optical transition integrals $\vec{M}_{a_1,a_2}(\tilde{\vec{A}})$ (Eq.~\eqref{eq:LMWW}) result in more non-vanishing transitions, effectively leading to the same reduction of the second selection rule as DCI. The third envelope selection rule is not touched by $\beta \neq 1$.

\underline{SRE:} SRE does not mix different envelopes, thus no envelope selection rule is affected. The spin basis states are chosen in a suitable basis for SRE, thus spin selection rules are also not affected. 

\underline{VBM:} If we just treat the VBM induced mixtures between envelopes, which follow the symmetry $\sum_{\alpha} (a_{\text{elec.}}^{\alpha} + a_{\text{hole}}^{\alpha}) - (\tilde{a}_{\text{elec.}}^{\alpha} + \tilde{a}_{\text{hole}}^{\alpha}) \in \{0, 2, 4, ... \}$, we see that the direction of excitation is not important any more and the pure envelope selection rules are reduced to $\text{modulo}(\Delta a^x+\Delta a^y+\Delta a^z,2)=\text{modulo}(n,2)$. Thus transitions with $\Delta a^z \neq 0$ become allowed and transitions previously just allowed with high $n$ become allowed already with lower $n$, like $d_{xy}\to s$ in $n=0$.
However, with VBM the coupling between two envelopes is always accompanied by a spin flip and the degree of freedom of spin and envelope get intermixed. Thus we should look for a combined spin and envelope selection rule. Such a rule can be deduced from Tab.~\ref{tab:subspaces}.

\section{Breaking of reflection symmetry}\label{sec:appsymmetrybreaking}

\begin{figure}[H]
\includegraphics[width=1.0\columnwidth]{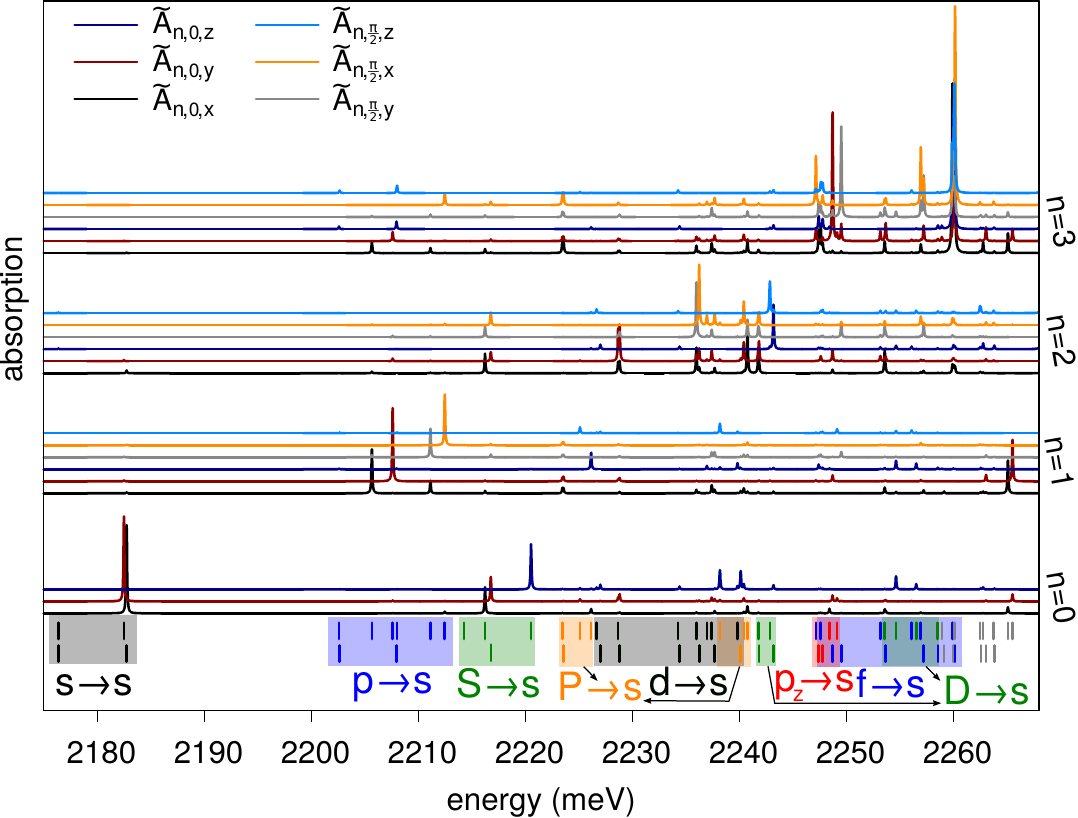}
\caption{Absorption spectra for a QD in our full model with broken reflection symmetry for different linear polarizations $\alpha$ and different rotations $\theta$. Below, all existing exciton eigenstates are displayed in blocks labeled by the most appropriate envelope basis state. Absorption lines and states have an independent coloring.}\label{fig:comparebrokenS}
\end{figure}

Our QD model preserves a threefold reflection symmetry (thus $D_{2h}$-symmetry), leading to eight subspaces which are separately addressable by light fields of the corresponding parity (see Tab.~\ref{tab:subspaces}). This last unbroken selection rule is not valid any more if we break the reflection symmetries.\\
One possibility for such a symmetry breaking is a more complex shape of the QD confinement. A typical example for a broken reflection symmetry in $z$-direction is a pyramidal QD ($C_{2v}$-symmetry). This would mix the spectra of light fields with even (odd) $n$ and $x/y$-polarization and those with odd (even) $n$ and $z$-polarization. For a mixture of the spectra of equally polarized light fields and just different parities of $n$, we need a broken reflection symmetry in in-plane direction. Therefore one would need for example a QD with a pear-shape in in-plane direction, what is a less commonly supposed geometry. In fact there are both, QDs with indications for a strongly broken \cite{benny2012excitation} and well preserved \cite{hinz2018charge} $C_{2v}$-symmetry. To estimate the influence of a broken reflection symmetry of the QD confinement, we consider a general potential $\sum_{\alpha \in \{x,y,z\}} \hbar \omega_{b,\alpha} \sum_n C^{\alpha}_n \, \left(\sqrt{\frac{m_{b,\alpha}\omega_{b,\alpha}}{2 \hbar}} \alpha \right)^n$ with the coefficients $C^{\alpha}_n$. We just study the first and second order terms with $C^{\alpha}_2=1$, $C^{\alpha}_1=\sqrt{2} C^{\alpha}$ and all other $C^{\alpha}_n=0$. This results in a displaced quadratic potential. When shifting the potential for electrons and holes in different directions, we break the reflection symmetry in the respective direction.
To break the reflection symmetry noticeably within the absorption spectra we set $C^x=C^y=C^z=\frac{1}{4}$, corresponding to a distance between the center of the electron and hole confinement of around $\frac{1}{4} L_{\alpha}$ \footnote{This is exact for $\beta=1$} in the respective direction. The resulting absorption spectra are displayed in Fig.~\ref{fig:comparebrokenS} and fit well to the theoretical discussion of a broken reflection symmetry in Sec.~\ref{sec:coulombinteractionvbm}. A typical effect is visible for the $(p_{\text{inpl.}}\to s) \epsilon_{x/y}$ transitions, which become slightly bright in the spectra of light modes with $n=0$ and $n=2$.

Also static electric fields $\vec{E}$ can break the reflection symmetry by the Hamiltonian $\hat{H}_{\text{electr.}}=-q\vec{E}\cdot \hat{\vec{r}}$ with opposite charges $q$ of the hole/electron. For plane-wave like excitation and our QD parameters, we would need field strengths of around 5~$\frac{\text{meV}}{\text{nm}}$ to get intensities of $p_x\to s$ similar to those of $d_{\text{inpl.}}\to s$. This could enable another way to excite for example $p\to s$ excitons by shortly activating an electric field while exciting with plane wave like light.

The selection rules can also be broken by a reduced symmetry of the light fields.

\section{Measurement of $d_{\text{inpl.}}\to s$ eigenstates}\label{sec:appds}

From theoretical considerations, one knows that the approximate eigenstates mostly consisting of $d_{\text{inpl.}}\to s$ are given via (from lower to higher energy)

\begin{figure}[H]
\includegraphics[width=1.0\columnwidth]{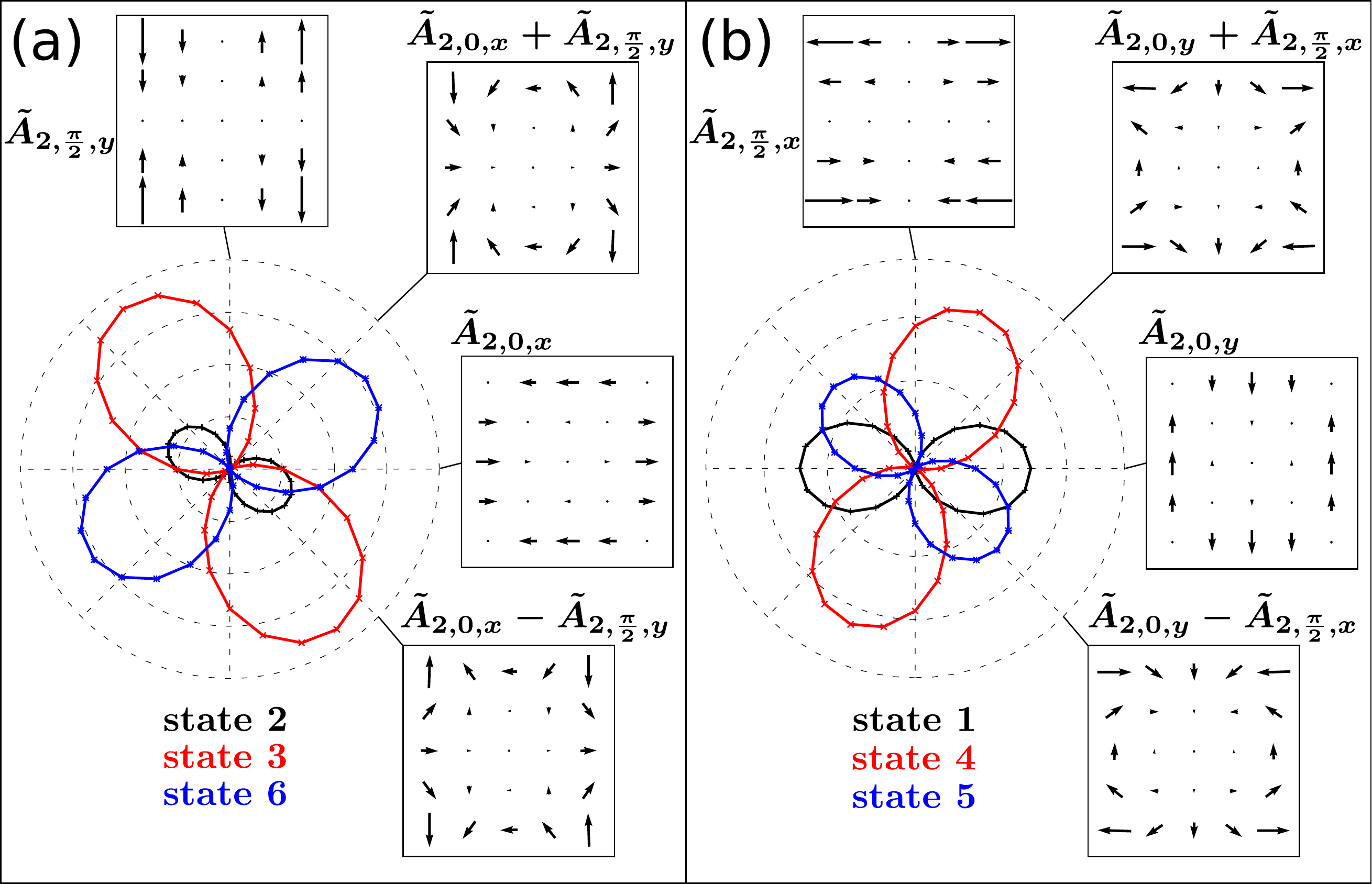}
\caption{Intensity of the $(d_{\text{inpl.}}\to s) \epsilon_{x/y}$ eigenstates for different superpositions of $\tilde{\vec{A}}_{2,0,x}$ and $\tilde{\vec{A}}_{2,\frac{\pi}{2},y}$ ($\tilde{\vec{A}}_{2,0,y}$ and $\tilde{\vec{A}}_{2,\frac{\pi}{2},x}$) in a (b). The insets display the corresponding field profiles.
}\label{fig:symmQDPolards}
\end{figure}

{\footnotesize
\begin{align}
\text{state ds1: } & \frac{1}{\sqrt{2}}[C_1(d_{xx}\to s)\epsilon_{y} + \tilde{C}_1(d_{yy}\to s)\epsilon_{y}] + \tilde{\tilde{C}}_1(d_{xy}\to s) \epsilon_{x}, \notag\\
\text{state ds2: } & \frac{1}{\sqrt{2}}[C_2(d_{xx}\to s) \epsilon_{x} + \tilde{C}_2(d_{yy}\to s) \epsilon_{x}] - \tilde{\tilde{C}}_2(d_{xy}\to s) \epsilon_{y}, \notag\\
\text{state ds3: } & C_3(d_{xy}\to s) \epsilon_{y} - \frac{1}{\sqrt{2}}[\tilde{C}_3(d_{xx}\to s)\epsilon_{x} - \tilde{\tilde{C}}_3(d_{yy}\to s)\epsilon_{x}], \notag\\
\text{state ds4: } & C_4(d_{xy}\to s) \epsilon_{x} + \frac{1}{\sqrt{2}}[\tilde{C}_4(d_{xx}\to s)\epsilon_{y} - \tilde{\tilde{C}}_4(d_{yy}\to s)\epsilon_{y}], \notag\\
\text{state ds5: } & \frac{1}{\sqrt{2}}[\tilde{C}_5(d_{xx}\to s)\epsilon_{y} - C_5(d_{yy}\to s)\epsilon_{y}]  - \tilde{\tilde{C}}_5(d_{xy}\to s) \epsilon_{x},  \notag\\
\text{state ds6: } & \frac{1}{\sqrt{2}}[\tilde{C}_6(d_{xx}\to s)\epsilon_{x} - C_6(d_{yy}\to s)\epsilon_{x}]  + \tilde{\tilde{C}}_6(d_{xy}\to s) \epsilon_{y}, \label{eq:eigenstd}
\end{align}
}

with typically $\tilde{\tilde{C}}_i > \tilde{C}_i$ and $C_i > \tilde{\tilde{C}}_i$.
The light modes $\tilde{\vec{A}}_{n=2,\theta=0,\alpha} \sim (x^2-y^2) \vec{e}_{\alpha}$ and $\tilde{\vec{A}}_{n=2,\theta=\frac{\pi}{2},\alpha} \sim xy \vec{e}_{\alpha}$ excite the states $(d_{xx}\to s) \epsilon_{\alpha}-(d_{yy}\to s) \epsilon_{\alpha}$ and $(d_{xy}\to s)\epsilon_{\alpha}$, respectively. The absorption for different real superpositions of these modes is given in Fig.~\ref{fig:symmQDPolards}.

With these two light modes, we can just reveal the coefficients of a 2-dimensional basis. A change from the basis states $(d_{xx}\to s)\epsilon_{\alpha}$ and $(d_{yy}\to s)\epsilon_{\alpha}$ to $(d_{xx}\to s)\epsilon_{\alpha} \pm (d_{yy}\to s)\epsilon_{\alpha}$, where $(d_{xx}\to s)\epsilon_{\alpha} + (d_{yy}\to s)\epsilon_{\alpha}$ is dark so its coefficient cannot be measured anyway, provides a well defined 2-dimensional basis.
The eigenstates obtained from the simulated measurement in Fig.~\ref{fig:symmQDPolards} fit well to the theoretically predicted form of Eqs.~\eqref{eq:eigenstd}.
The angle mismatch between the ``measured'' eigenstate and the ``exact'' eigenstate without higher terms is 3\degree{}-14\degree{}. The accuracy is reduced compared to the measurement of the $p_{\text{inpl.}}\to s$ states, since the $d_{\text{inpl.}}\to s$ eigenstates have larger contributions of energetically higher basis states (more than $30\%$).

\section{Different representations of Bessel beams}\label{sec:appBesToNod}
An exact solution of the Helmhotz equation is given by Bessel beams \cite{volke2002orbital, quinteiro2014light}. We discuss three equivalent representations, as listed below. For all cases, we give approximations up to a useful level. Therefore we consider $\frac{q_r}{q_z} \approx 1$ and a region around $q_r r=0$, thus $ J_n(q_r r) = \sum_{j=0}^{\infty} \frac{(-1)^j (\frac{q_r r}{2})^{2j+n}}{(j+n)!j!} \approx \frac{(q_r r)^n}{2^n n!}$ and $(q_r r)^{n+1} \ll (q_r r)^{n}$. One can construct standing waves by a superposition of two waves propagating in opposite directions. Thereby different local fields arise at different values of $z$. For small $q_z z$ we use $\cos(q_z z) \approx 1$ and $\sin(q_z z) \approx q_z z$. We use the cylindrical coordinates $r$, $\varphi$ and $z$.\\

\begin{widetext}

\underline{\textbf{1. Twisted light beams}}\\
This representation is similar to a complex-valued cylindrical multipole expansion, Laguerre-Gaussian beams, so called vortex beams or twisted light. The fields can be described by nodal planes with a normal in in-plane direction, which rotate in space and time, giving them the nickname ``twisted light''. Within the paraxial limit, the indices $l$ and $\sigma$ label the orbital angular momentum and circular polarization of the mode, respectively.

\textbf{Traveling waves}
\begin{align}
\vec{A}_{l,\sigma}(\vec{r},t) & = A_{0} e^{i (q_{z} z-\omega t)}
\left[ J_{l}(q_{r} r) e^{i l \varphi} \frac{1}{\sqrt{2}} \left(\vec{e}_{x}+i\sigma \vec{e}_{y} \right) - \frac{i \sigma}{\sqrt{2}} 
\frac{q_{r}}{q_{z}} J_{l+\sigma}(q_{r} r) e^{i (l+\sigma) \varphi} \vec{e}_{z} \right] + c.c. \notag\\
& \approx \begin{cases}
A_{0} e^{i (q_{z} z-\omega t)}
\left[ J_{l}(q_{r} r) e^{i l \varphi} \frac{1}{\sqrt{2}}\left(\vec{e}_{x}+i\sigma \vec{e}_{y} \right)  \right] + c.c. \, , \text{for } \text{sign}(l)=\text{sign}(\sigma) \\
A_{0} e^{i (q_{z} z-\omega t)}
\left[ - \frac{i \sigma}{\sqrt{2}} 
\frac{q_{r}}{q_{z}} J_{l+\sigma}(q_{r} r) e^{i (l+\sigma) \varphi} \vec{e}_{z} \right] + c.c. \, , \text{for } \text{sign}(l)\neq \text{sign}(\sigma) \\
\end{cases} \label{eq:tlEfield}
\end{align}

\underline{\textbf{2. Radially and azimuthally polarized beams}}\\
This representation is somewhere between the complex- and real-valued cylindrical multipole expansions. Radially and azimuthally polarized beams are exemplary realizations. These modes become interesting for QDs with cylindrical symmetry or fields with a strong component polarized in propagation direction (here $z$-direction). Corresponding standing waves are given in Eqs.~\eqref{eq:Bessel_xy_z_standing}.

\textbf{Traveling waves}
\begin{align}
\vec{A}^{xy}_{n,\theta}(\vec{r},t) & = 
\begin{cases}
\frac{ 1}{\sqrt{2}} [\vec{A}_{n,1}(\vec{r},t) + (-1)^n \vec{A}_{-n,-1}(\vec{r},t) ] \, , \text{for } \theta=0 \\
\frac{-i}{\sqrt{2}} [\vec{A}_{n,1}(\vec{r},t) - (-1)^n \vec{A}_{-n,-1}(\vec{r},t) ] \, , \text{for } \theta=\frac{\pi}{2} \\
\end{cases} \notag\\
& = A_{0} e^{i (q_{z} z-\omega t)} \left[ J_{n}(q_{r} r) \left( \cos(n \varphi-\theta) \vec{e}_x - \sin(n \varphi - \theta) \vec{e}_y \right) - i \frac{q_r}{q_z} J_{n+1}(q_{r} r) \cos((n+1) \varphi-\theta) \vec{e}_z \right] + c.c. \notag\\
& \approx A_{0} 2 \cos(q_{z} z-\omega t) \frac{q_r^n r^n}{2^n n!} \left( \cos(n \varphi-\theta) \vec{e}_x - \sin(n \varphi - \theta) \vec{e}_y \right) \notag\\
\vec{A}^{z}_{n,\theta}(\vec{r},t) & = 
\begin{cases}
\frac{ 1}{\sqrt{2}} [\vec{A}_{n+1,-1}(\vec{r},t) + (-1)^{n+1} \vec{A}_{-(n+1),1}(\vec{r},t) ] \, , \text{for } \theta=0 \\
\frac{-i}{\sqrt{2}} [\vec{A}_{n+1,-1}(\vec{r},t) - (-1)^{n+1} \vec{A}_{-(n+1),1}(\vec{r},t) ] \, , \text{for } \theta=\frac{\pi}{2} \\
\end{cases} \notag\\
& = A_{0} e^{i (q_{z} z-\omega t)} \left[ J_{n+1}(q_{r} r) \left( \cos((n+1) \varphi-\theta) \vec{e}_x + \sin((n+1) \varphi - \theta) \vec{e}_y \right) + i \frac{q_r}{q_z} J_{n}(q_{r} r) \cos(n \varphi-\theta) \vec{e}_z \right] + c.c. \notag\\
& \approx
\begin{cases}
A_{0} 2 \cos(q_{z} z-\omega t) \frac{q_r r}{2} \left( \cos(\varphi-\theta) \vec{e}_x + \sin(\varphi - \theta) \vec{e}_y \right) \, , \text{for } n=0; \theta=\frac{\pi}{2} \\
A_{0} 2 \sin(q_{z} z-\omega t) \frac{q_r}{q_z} \frac{q_r^{n} r^{n}}{2^{n} n!} \cos(n \varphi-\theta) \vec{e}_z \, , \text{else}
\end{cases} \label{eq:Bessel_xy_z}
\end{align}

\textbf{Standing waves}
\begin{align}
\vec{\breve{A}}^{xy}_{n,\theta}(\vec{r},t) & = A_{0} 2 \cos(\omega t) \left[ \cos(q_{z} z) J_{n}(q_{r} r) \left( \cos(n \varphi-\theta) \vec{e}_x - \sin(n \varphi - \theta) \vec{e}_y \right) + \sin(q_{z} z) \frac{q_r}{q_z} J_{n+1}(q_{r} r) \cos((n+1) \varphi-\theta) \vec{e}_z \right] \notag\\
& \approx \frac{2 A_{0}}{2^n n!} \cos(\omega t) \left[ q_r^n r^n \left( \cos(n \varphi-\theta) \vec{e}_x - \sin(n \varphi - \theta) \vec{e}_y \right) +  \frac{1}{2 n (n+1)} \frac{n z}{r} q_r^{n+2} r^{n+2} \cos((n+1) \varphi-\theta) \vec{e}_z \right] \notag\\
%
\vec{\breve{A}}^{z}_{n,\theta}(\vec{r},t) & = A_{0} 2 \cos(\omega t) \left[ \cos(q_{z} z) J_{n+1}(q_{r} r) \left( \cos((n+1) \varphi-\theta) \vec{e}_x + \sin((n+1) \varphi - \theta) \vec{e}_y \right) \phantom{\frac{1}{1}} \right. \notag\\
&\left. \quad\quad\quad\quad\quad\quad\quad\quad\quad\quad\quad\quad\quad   - \sin(q_{z} z) \frac{q_r}{q_z} J_{n}(q_{r} r) \cos(n \varphi-\theta) \vec{e}_z \right] \notag\\
& \overbrace{\approx}^{z \approx 0} \frac{2 A_{0}}{2^{n+1} (n+1)!} \cos(\omega t) \left[ q_r^{n+1} r^{n+1} \left( \cos((n+1) \varphi-\theta) \vec{e}_x + \sin((n+1) \varphi - \theta) \vec{e}_y \right) \phantom{\frac{1}{1}} \right. \notag\\
&\left. \quad\quad\quad\quad\quad\quad\quad\quad\quad\quad\quad\quad\quad - 2 \frac{(n+1) z}{r} q_r^{n+1} r^{n+1} \cos(n \varphi-\theta) \vec{e}_z \right] \notag\\
& \text{or} \notag\\
& \overbrace{\approx}^{z \approx -\frac{\pi}{2 q_z}} \frac{q_r}{q_z} \frac{2 A_{0}}{2^{n} n!} \cos(\omega t) \left[ \frac{1}{2 (n+1)} \frac{1}{(\frac{q_r}{q_z})^2} \frac{z+\frac{\pi}{2 q_z}}{r} (q_r r)^{n+2} \left( \cos((n+1) \varphi-\theta) \vec{e}_x + \sin((n+1) \varphi - \theta) \vec{e}_y \right) \right. \notag\\
&\left. \phantom{\frac{1}{1}} \quad\quad\quad\quad\quad\quad\quad\quad\quad\quad\quad\quad\quad + (q_{r} r)^{n} \cos(n \varphi-\theta) \vec{e}_z \right] \notag\\
& \approx \frac{q_r}{q_z} 2 A_{0} \cos(\omega t) \frac{(q_{r} r)^{n}}{2^{n} n!} \cos(n \varphi-\theta) \vec{e}_z \label{eq:Bessel_xy_z_standing}
\end{align}

\underline{\textbf{3. Real-valued cylindrical multipole modes}}\\
This representation is similar to a real-valued cylindrical multipole expansion, as used throughout this paper. Compared to the complex-valued expansion, the nodal planes of the field have a fixed orientation in space and time.

\textbf{Traveling waves}
\begin{align}
& \vec{A}^{x}_{n,\theta}(\vec{r},t) = 
\begin{cases}
\vec{A}^{xy}_{0,0}(\vec{r},t) \, , \text{for } n=0; \theta=0 \\
\frac{1}{2} [\vec{A}^{xy}_{n,\theta}(\vec{r},t) + \vec{A}^{z}_{n-1,\theta}(\vec{r},t)] \, , \text{for } n\geq 1
\end{cases} \notag\\
& = A_{0} e^{i (q_{z} z-\omega t)} \left[ J_{n}(q_{r} r) \cos(n \varphi-\theta) \vec{e}_x - \frac{i}{2} \frac{q_r}{q_z} \lbrace J_{n+1}(q_{r} r) \cos((n+1) \varphi-\theta) - J_{n-1}(q_{r} r) \cos((n-1) \varphi-\theta)  \rbrace \vec{e}_z \right] + c.c. \notag\\
& \vec{A}^{y}_{n,\theta}(\vec{r},t) = 
\begin{cases}
\vec{A}^{xy}_{0,\frac{\pi}{2}}(\vec{r},t) \, , \text{for } n=0,\theta=0 \\
\frac{1}{2} [\vec{A}^{xy}_{n,\frac{\pi}{2}}(\vec{r},t) - \vec{A}^{z}_{n-1,\frac{\pi}{2}}(\vec{r},t)] \, , \text{for } n\geq 1,\theta=0 \\
\frac{-1}{2} [\vec{A}^{xy}_{n,0}(\vec{r},t) - \vec{A}^{z}_{n-1,0}(\vec{r},t)] \, , \text{for } n\geq 1,\theta=\frac{\pi}{2}
\end{cases} \notag\\
& = A_{0} e^{i (q_{z} z-\omega t)} \left[ J_{n}(q_{r} r) \cos(n \varphi-\theta) \vec{e}_y - \frac{i}{2} \frac{q_r}{q_z} \lbrace J_{n+1}(q_{r} r) \sin((n+1) \varphi-\theta) + J_{n-1}(q_{r} r) \sin((n-1) \varphi-\theta)  \rbrace \vec{e}_z \right] + c.c. \label{eq:Bessel_x_y}
\end{align}

\textbf{Standing waves}
\begin{align}
& \vec{\breve{A}}^{x}_{n,\theta}(\vec{r},t) = A_{0} 2 \cos(\omega t) \cdot \notag\\
& \cdot \left[ \cos(q_{z} z) J_{n}(q_{r} r) \cos(n \varphi-\theta) \vec{e}_x + \frac{1}{2} \sin(q_{z} z) \frac{q_r}{q_z} \lbrace J_{n+1}(q_{r} r) \cos((n+1) \varphi-\theta) - J_{n-1}(q_{r} r) \cos((n-1) \varphi-\theta)  \rbrace \vec{e}_z \right] \notag\\
& \approx \frac{2 A_{0}}{2^n n!} \cos(\omega t) \left[ q_r^n r^n \cos(n \varphi-\theta) \vec{e}_x - \frac{n z}{r} q_r^n r^{n} \cos((n-1) \varphi-\theta) \vec{e}_z \right] \notag\\
& \vec{\breve{A}}^{y}_{n,\theta}(\vec{r},t) = A_{0} 2 \cos(\omega t) \cdot \notag\\
& \cdot \left[ \cos(q_{z} z) J_{n}(q_{r} r) \cos(n \varphi-\theta) \vec{e}_y + \frac{1}{2} \sin(q_{z} z) \frac{q_r}{q_z} \lbrace J_{n+1}(q_{r} r) \sin((n+1) \varphi-\theta) + J_{n-1}(q_{r} r) \sin((n-1) \varphi-\theta)  \rbrace \vec{e}_z \right] \notag\\
& \approx \frac{2 A_{0}}{2^n n!} \cos(\omega t) \left[ q_r^n r^n \cos(n \varphi-\theta) \vec{e}_y + \frac{n z}{r} q_r^n r^{n} \sin((n-1) \varphi-\theta) \vec{e}_z \right] \label{eq:Bessel_x_y_standing_approx}
\end{align}
\end{widetext}


%

\end{document}